\documentstyle[12pt,a4wide]{article}
\newcommand{\be}{\begin{equation}}
\newcommand{\ee}{\end{equation}}
\newcommand{\bea}{\begin{eqnarray}}
\newcommand{\eea}{\end{eqnarray}}
\newcommand{\bean}{\begin{eqnarray*}}

\newcommand{\eean}{\end{eqnarray*}}
\font\upright=cmu10 scaled\magstep1
\font\sans=cmss10
\newcommand{\ssf}{\sans}
\newcommand{\stroke}{\vrule height8pt width0.4pt depth-0.1pt}
\newcommand{\Z}{\hbox{\upright\rlap{\ssf Z}\kern 2.7pt {\ssf Z}}}

\newcommand{\C}{{\rlap{\rlap{C}\kern 3.8pt\stroke}\phantom{C}}}
\newcommand{\R}{\hbox{\upright\rlap{I}\kern 1.7pt R}}
\newcommand{\CP}{\C{\upright\rlap{I}\kern 1.5pt P}}
\newcommand{\PP}{\hbox{\upright\rlap{I}\kern 1.5pt P}}

\newcommand{\identity}{{\upright\rlap{1}\kern 2.0pt 1}}

\newcommand{\HH}{\mbox{\hbox{\upright\rlap{I}\kern 1.7pt H}}}

\newcommand{\vp}{\vert\phi\vert}

\font\mybb=msbm10 at 11pt

\def\bb#1{\hbox{\mybb#1}}

\def\bC {\bb{C}}

\renewcommand{\CP}{\bC {\rm P}}

\newcommand{\news}{\setcounter{equation}{0}}
\input{epsf}
\begin{document}
\title{\vskip -70pt
\begin{flushright}
\end{flushright}\vskip 50pt
{\bf \LARGE \bf Q-ball Dynamics }\\[30pt]
\author{Richard A. Battye$^{\ \dagger}$ and Paul M. Sutcliffe$^{\ \ddagger}$
\\[10pt]
\\{\normalsize $\dagger$ {\sl Department of Applied Mathematics and Theoretical
Physics,}}
\\{\normalsize {\sl Centre for Mathematical Sciences, University of Cambridge,}}
\\{\normalsize {\sl Wilberforce Road, Cambridge CB3 0WA, U.K.}}
\\{\normalsize {\sl Email : R.A.Battye@damtp.cam.ac.uk}}\\
\\{\normalsize $\ddagger$  {\sl Institute of Mathematics, University of Kent at Canterbury,}}\\
{\normalsize {\sl Canterbury, CT2 7NZ, U.K.}}\\
{\normalsize{\sl Email : P.M.Sutcliffe@ukc.ac.uk}}\\}}
\date{March 2000}
\maketitle

\begin{abstract}
We investigate the dynamics of Q-balls in one, two and three space dimensions,
using numerical simulations of the full nonlinear equations of motion.
We find that the dynamics of Q-balls is extremely complex, involving
processes such as charge transfer and Q-ball fission.
 We present results of simulations which illustrate the salient features of 2-Q-ball interactions
 and give qualitative arguments to explain them in terms of the evolution of the time-dependent phases.
\end{abstract}

\newpage
\section{Introduction}
\news\ \ \ \ \ \
One of the most fascinating areas of inter-disciplinary
research at the interface between mathematics and physics is the study
of {\it solitons}. This word has as many definitions as there are
people who study them, but in general terms they are stable, localized energy distributions. From a
purely mathematical perspective, solitons are described as
extended solutions to a set of hyperbolic or parabolic partial differential
equations, which can travel without dissipation at a uniform velocity
and maintain, at least asymptotically, their shape during collisions;
often these properties of solitons are attributable to the existence of 
an infinite number of conserved quantities, connected with the notions integrability,
and radiation-free soliton collisions can be constructed.

In the context of particle physics, which is our main interest here,
 the usage of the word soliton is less rigorous and any kind of
 localized energy distribution falls under this broad umbrella. Only
 in very specialised circumstances will soliton collisions not generate
radiation and solutions which radiate substantially during, for
 example, a highly relativistic 2-soliton collision are included. Of course 
 very little exact analytic progress is possible in these more general
settings since radiative processes are
 notoriously difficult to model analytically, thus numerical simulations
 are necessary to probe more complicated situations. The usual
 development of  understanding in this subject follows an intricate
 interplay between the two, with analytic work putting on solid ground
 more qualitative observations from simulations. The domain of
 validity of analytic approximations can then be checked and extended
 by further simulations. Examples of the classes of solitons which
 have been investigated in this way are vortices~\cite{vort}, monopoles~\cite{mon} and skyrmions~\cite{sky}, 
whose existence and stability is essentially due to conserved topological currents and charges, 
along with energy bounds related to the charge and stable scaling laws.
In these examples the topological features constrain the amount of radiation produced
in low energy collisions and allows approximations to be applied which ignore the
radiative effects.

The subject of this paper is a particular class of solitons known as
Q-balls~\cite{lee,coleman}. These are different in many ways to the
topological solitons mentioned above. Firstly, they are time-dependent
with a rotating internal phase. Secondly, the conserved charge
associated with their stability, Noether charge (Q), is
not topological and therefore their stability is also a dynamical issue.
 As we will see these two features lead to a much more complicated variety of interaction properties
 than seen in the study of topological solitons. The main difference is that the charge is quantized in
 topological models, it usually being scaled to be an exact integer, whereas we shall see that the  charge Q 
can take any value (in a specified range) allowing for the possibility of charge transfer between solitons 
 and/or fission during the interaction process.

Although the concepts associated with Q-balls are extremely general and
they are likely to exist in a wide variety of physical contexts (for
example, see ref.~\cite{lowtemp}), the main motivation for the current
study is the recent realization that they are a generic consequence of
the Minimal Supersymmetric Standard Model (MSSM)~\cite{k2} due to
the existence of D-flat directions in the effective potential created
by tri-linear couplings. In
this context the conserved charge is that associated with the U(1)
symmetries of Baryon and Lepton number conservation and the relevant U(1)
fields correspond to either squark or slepton particles. Therefore, the
Q-balls can be thought of as  condensates of either a large number of squarks or
sleptons. It has been suggested that such condensates can be involved
in baryogenesis via the Affleck-Dine mechanism~\cite{AD} after an
epoch of inflation in the early universe.
If this is the case then there are two interesting
possibilities. If the Q-balls can avoid evaporation into lighter,
stable  particles such as  protons, then it might be possible for them
to be important cosmologically as cold dark matter~\cite{KS}. Whereas
if they are unstable, they would decay in a non-trivial way
into baryons and could  lead
to observable isocurvature baryon fluctuations~\cite{EM}.

Underpinning these interesting suggestions are assumptions as to how
Q-balls actually interact and it is our intention here to make an
exhaustive study of this issue. Our approach will be to identify
numerically the important dynamical processes that can occur in
general situations of two interacting Q-balls, which we will then
explain qualitatively, leaving a more detailed analytic exposition of
the dynamics~\cite{BMS} and the cosmological implications to
subsequent papers. In the next section we will discuss the basic
properties of static U(1) Q-balls. Then we will present a detailed and
extensive study of Q-balls on the line. As we will see, many of the
properties of interest such as fission and charge transfer are
observed in one-dimension and given the simplicity of simulations, it
seems sensible to make the most exhaustive study there. In the
subsequent sections on planar Q-balls and fully three-dimensional
Q-balls we will show to what extent the one-dimensional simulations can
be used to understand the dynamics in higher dimensions and what
effects are clearly of higher dimensional origin. In a penultimate section
we will discuss the interactions of Q-balls with anti-Q-balls which have
an equal and opposite rotation, before making a concluding summary
in the final section.

We should note that there is a disparate literature~\cite{lit} on
Q-balls in which some (but by no means all) of the processes we will discuss  have already
been noted, but not necessarily completely understood. In particular
we should mention recent work~\cite{greek} which presented results for
Q-balls in one and two dimensions. There it was suggested that the
right-angle scattering of solitons seen in two-dimensional
topological soliton models is also prevalent in these non-topological
models. At the relevant points we will point out in what ways we disagree
with their explanation of this phenomena, and demonstrate that it is by no means general.

\section{U(1) Q-balls}
\news\ \ \ \ \ \

Given our motivation for studying Q-balls it seems sensible
to work with the U(1) Goldstone model, although Q-balls can exist in a
variety of field theoretic models. To be precise, the model we consider is that of a single complex scalar
field $\phi$ in $D=1,2,3$ spatial dimensions with a potential $U(\vp).$
Explicitly, the Lagrangian is
\be{\cal L}=\frac{1}{2}\partial_\mu\phi \partial^\mu\bar\phi-
U(\vp)\,,\label{lag}\ee
with the key feature being the fact that the potential is only a
function of $\vp$.
The model has a global $U(1)$ symmetry
and the associated conserved Noether current $J_{\mu}$ is given by 
\be
J_{\mu}={1\over
2i}\left(\bar\phi\partial_{\mu}\phi-\phi\partial_\mu\bar\phi\right)\,,
\ee
whose covariant conservation $\partial^{\mu}J_{\mu}=0$ leads to the
existence of the conserved Noether charge Q, given by
\be 
Q={1\over 2i}\int\left(\bar\phi\partial_t\phi-\phi\partial_t\bar\phi\right)d^Dx=\int \mbox{Im}(\bar\phi \partial_t\phi) \ d^Dx\,.
\label{q}\ee
A stationary Q-ball solution has the form
\be
\phi=e^{i\omega t}f(r)\,,
\label{qbform}\ee
where  $f(r)$ is a real
radial profile function which satisfies the ordinary differential equation
\be
\frac{d^2f}{dr^2}=\frac{(1-D)}{r}\frac{df}{dr}-\omega^2 f+U'(f)\,,
\label{profile}\ee
with the boundary conditions that $f(\infty)=0$ and $\frac{df}{dr}(0)=0.$

This equation can either be interpreted as describing the motion of a
point particle moving in a potential with friction~\cite{coleman}, or
in terms of Euclidean bounce solutions~\cite{k1}; in
each case the effective potential being $U_{\rm
eff}(f)=\omega^2f^2/2-U(f)$. This  leads to constraints
on the potential $U(f)$ and the frequency $\omega$ in order for a Q-ball solution to
exist. Firstly, the effective mass of $f$ must be negative. If we
consider a potential $U(f)$ which is non-negative and satisfies
$U(0)=U'(0)=0$, $U''(0)=\omega_+^2>0$, then one can deduce that
$\omega<\omega_+$. Furthermore, the minimum of $U(f)/f^2$ must be
attained at some positive value of $f$, say $0<f_0<\infty,$ and existence of
the solution requires that $\omega>\omega_-$ where 
\be \omega_-^2=2U(f_0)/f_0^2\,.
\label{range}\ee
Hence,  Q-ball solutions exist for all $\omega$ in the range
$\omega_-<|\omega|<\omega_+$. Note (i) that solutions exist for
positive and negative values of $\omega$, the negative ones being
termed anti-Q-balls, (ii) it is often interesting to think of the
Q-balls as being akin to charged bubbles; their profiles being very similar. 

The classical stability of the
solutions is a more sensitive issue. For sufficiently large $Q$ these
solutions are guaranteed to be stable, as can be seen using the \lq thin wall
limit\rq\ \cite{coleman}, where the profile function can be
approximated by a smoothed-out step function. For small Q it is
necessary to perform a full stability analysis using the second
variation of the action. In general, the results depend on the details of the
potential, but it can be shown that arbitrarily small Q-balls are
stable for certain potentials~\cite{k1}. For a rigorous approach to the
classical stability of Q-balls see ref.~\cite{St}
and references therein. From the quantum mechanical
point of view, the solutions are always stable for large enough $Q$ since the energy per
unit charge approaches $\omega_-$, which is always less than that for the $\phi$ particle itself, which
is $\omega_{+}$.

In choosing a simple potential which admits Q-balls there are
 three natural classes  which have been considered, although there are
 obviously many other possibilities,
\bea
&I&: \ \ U(f)=\alpha_1f^2+\alpha_2f^4+\alpha_3f^6\,,\\
&II&: \ \ U(f)=\beta_1f^2+\beta_2f^3+\beta_3f^4\,,\\
&III&: \ \ U(f)=\gamma_1f^2(1-\gamma_2\log(\gamma_3 f^2))+\gamma_4f^{2p}\,.
\eea
In each case it is possible to remove two of the parameters by
 rescaling the units of energy and time. Note therefore that the
 potentials of type I and II have one free parameter, while potentials
 of type III have two free parameters for a fixed value of $p$.

The type I potential is the simplest allowed potential which is a polynomial
in $f^2$, while type II is the simplest  which is polynomial in
$f$. Finally, those of type III  mimic the D-flat
direction in the MSSM. Here $p\ge 6$ is some integer that ensures the 
growth of the potential for large $f$, but does not destroy the
flatness property for intermediate values of $f.$ We should note that
none of these types of potential are the kind which might be
associated with a renormalizable quantum field theory, but are typical of
effective theories incorporating radiative or finite temperature
corrections to a bare potential.

In this paper we will mainly be concerned with the type I potential,
although we have also studied the type II case. We should note
that although the existence and stability properties of Q-balls with
these potentials are somewhat different, the qualitative features of the 
dynamics appears to be almost independent of the potential. The reason for this is that the
main interaction processes that we will describe, charge transfer  and
fission, are due mainly to the time-dependent nature of the solution,
rather than the precise profile function.

Explicitly, we shall choose our potential so that
\be
U(f)=f^2(1+(1-f^2)^2)\,,
\label{pot}
\ee
and therefore in terms of the earlier notation we have that
$w_+=2$ and $w_-=\sqrt{2}$, so that stable Q-balls exist
for $\sqrt{2}<\omega<2$. To illustrate the important features of Q-ball
solutions we shall focus on the case of one dimension where the profile function equation (\ref{profile})
can be solved exactly~\cite{lee} to give
\be
f_\omega(r)=\sqrt{\frac{4-\omega^2}{2+\sqrt{2\omega^2-4}
\cosh(2r\sqrt{4-\omega^2})}}\,.
\label{pro1d}
\ee
The associated energy $E_\omega$ and charge $Q_\omega$ can then be
computed to be 
\be
Q_\omega=\sqrt{2}\omega\tanh^{-1}\left(\frac{2-\sqrt{2\omega^2-4}}
{\sqrt{2}\sqrt{4-\omega^2}}\right)
\,,\quad E_\omega=\frac{\sqrt{4-\omega^2}}{2}
+\frac{(\omega^2+2)}{2\omega}Q_\omega.
\label{qe1d}\ee

\begin{figure}[ht]
\begin{center}
\leavevmode
\epsfxsize=12cm\epsffile{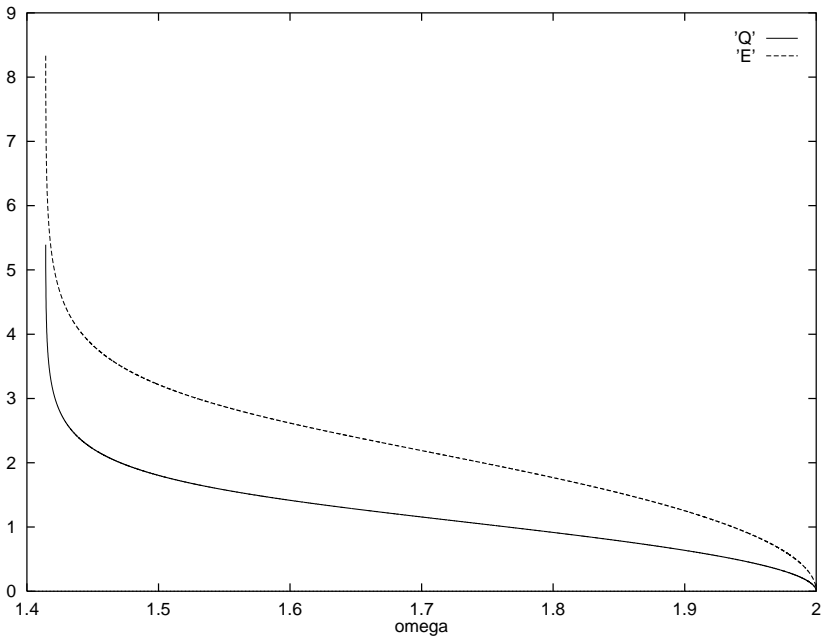}
\caption{The charge $Q$ and energy $E$ as a function of the frequency
$\omega.$ } 
\label{qvw}
\end{center}
\end{figure}

\begin{figure}[ht]
\begin{center}
\leavevmode
\epsfxsize=12cm\epsffile{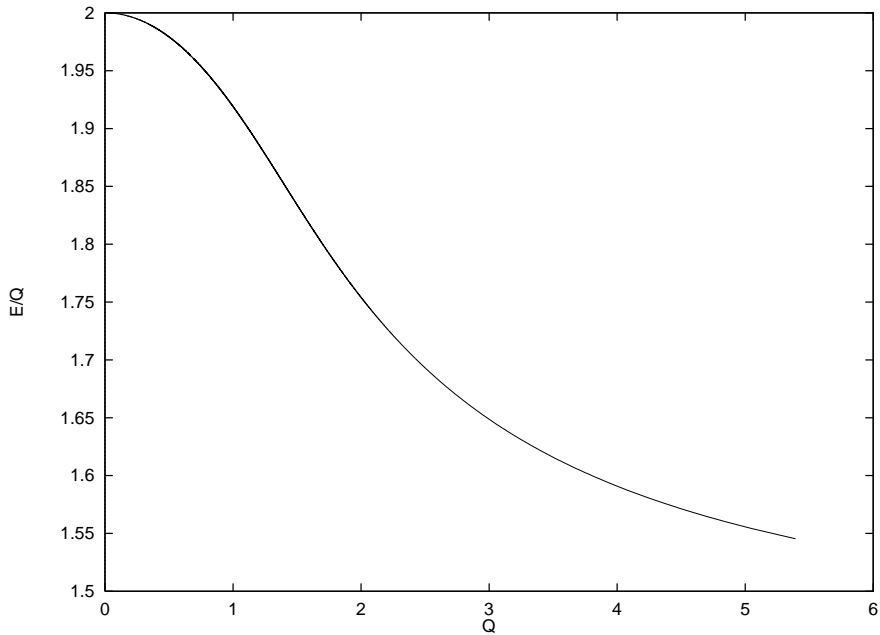}
\caption{The energy per unit charge $E/Q$ as a 
function of the charge $Q.$ }
\label{evq}
\end{center}
\end{figure}

In figure~\ref{qvw} we plot the charge $Q_\omega$ and energy
$E_\omega$ for $\omega$ in the allowed range  $\sqrt{2}<\omega<2.$
and we see that both $Q_\omega$ and  $E_\omega$  are monotonically
decreasing functions of $\omega.$ From (\ref{pro1d}) we can deduce that
\be
f_\omega(0)=\sqrt{{(4-\omega^2)}/{(2+\sqrt{2\omega^2-4)}}}\,,
\ee
and therefore 
$f_\omega(0)$ increases with the charge $Q_\omega$, since it is a decreasing
function of $\omega.$ In figure~\ref{evq} we display the energy per unit
charge $E_\omega/Q_\omega$ as a function of the charge $Q_\omega$, 
from which it can be seen that $E_\omega/Q_\omega$ is a decreasing
 function of the charge.
Recall that the asymptotic limit as $Q_\omega\rightarrow\infty$ is
$E_\omega/Q_\omega=\omega_-=\sqrt{2}.$  
Thus, these Q-balls are stable against decay into a number of smaller Q-balls
preserving the total charge.

Most of the general properties of Q-balls in any dimension
and for differing choices of the potential are captured by this
 one-dimensional example, where explicit formulae are available.
However, there are some slight differences, for example
if $D=3$ then there is a lower bound on the charge of a Q-ball,
whereas in the $D=1$ case considered above Q-balls can have an arbitrarily
small charge. This is not a generic feature of every potential and,
for example, it has been shown that for $D=3$ arbitrarily small Q-balls can be found
 with a potential of type II using a `thick
wall limit' in ref.~\cite{k1}. These kind of  
details are easily determined by solving the profile function
equation (\ref{profile}) numerically and a complete treatment of these
issues can be found in ref.~\cite{MV}. But, as we have already noted, we
don't believe that they are important for the dynamical processes 
which we focus on in the subsequent sections.

\vfill\eject

\section{Q-ball dynamics on the line}\news\ \ \ \ \ \

\noindent In this section we shall investigate the dynamics of 
Q-balls in one-dimension;  even for $D=1$ we shall see that
multi-Q-ball dynamics is a complicated issue, there being a rich variety of
phenomena associated with the non-quantization of the charge and the
time-dependent phase.

To investigate the dynamics of Q-balls we numerically solve the field
equations which follow from the Lagrangian (\ref{lag}) with
the potential (\ref{pot}), namely
\be
\ddot\phi-\nabla^2\phi+2\phi(2-4\vp^2+3\vp^4)=0\,,
\label{eom}
\ee
which is valid for any value of D.
The numerical methods we use are simple finite difference
schemes involving either second or fourth order accurate
spatial derivatives and a second order leapfrog algorithm
for the time evolution with 1000 points, the spatial step size $\Delta x=0.1$ 
and the time step size $\Delta t=0.02$. We apply absorbing boundary conditions,
which allows radiation to leave the grid and therefore simulates an
infinite domain (see refs.~\cite{BS,Bat} for details on
how to apply these boundary conditions).

As initial conditions to describe two well-separated Q-balls
we use the ansatz
\be
\phi=e^{i\omega_1t+i\alpha}f_{\omega_1}(\vert x+a\vert)
+e^{i\omega_2t}f_{\omega_2}(\vert x-a\vert )\,,
\label{ansatz1}
\ee
in one dimension, which can be trivially generalized to higher dimensions.
This ansatz describes a Q-ball with frequency $\omega_1$
at the position $x=-a$ and a second Q-ball with frequency
$\omega_2$ at the position $x=a.$ The $U(1)$ symmetry
of the theory means that for a single Q-ball the phase
of $\phi$ can be set to zero at $t=0$ without loss
of generality. However, for multi-Q-ball configurations only the
initial overall phase can be removed and so for a 2-Q-ball
configuration the relative phase, $\alpha$, remains as an important
parameter. 

The total charge of this configuration is 
\be
Q=Q_{\omega_1}+Q_{\omega_2}
+(\omega_1+\omega_2)\cos\alpha
\int_{-\infty}^\infty f_{\omega_1}(\vert x+a\vert)
f_{\omega_2}(\vert x-a\vert)\ dx.
\label{totalq}
\ee
The final term in the above expression is exponentially small
in the separation parameter $a$, since the profile functions have an 
exponential fall-off. However, we see that the relative phase $\alpha$ 
does affect the value of the total charge. Thus, strictly speaking, it
is not valid to substitute this ansatz into the energy functional to
determine how the energy depends on the relative phase, since this would
involve comparing configurations with differing values of $Q.$
The same remark also applies to any attempt to determine how the
potential energy depends upon the relative separation $a,$ which would
help to determine the nature of the interaction force between
Q-balls. This issue and its resolution using the methods of
ref.~\cite{ams} will be discussed in ref.~\cite{BMS}.
\begin{figure}[ht]
\begin{center}
\leavevmode
\epsfxsize=12cm\epsffile{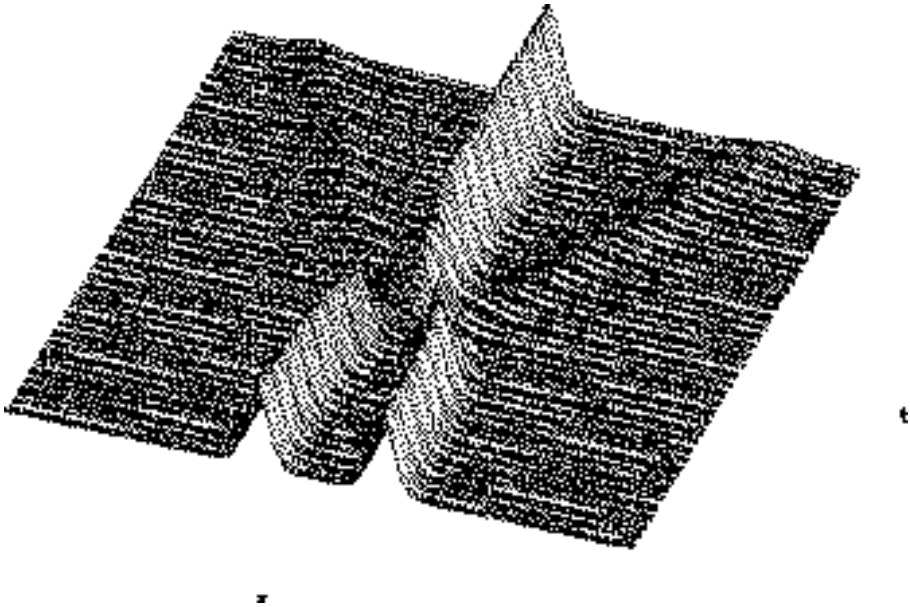}
\caption{The charge density at times $0\le t\le 300$,
for the parameters $\omega_1=\omega_2=1.5$, $a=4$, $\alpha=0.$ The two
Q-balls form a much larger Q-ball, almost stationary at the origin, and
the excess is taken away by the fission products: two small Q-balls.} 
\label{inphase}
\end{center}
\end{figure}

To begin with we consider configurations in which the two Q-balls have
the same charge, that is $\omega_1=\omega_2=\omega.$  In figure
\ref{inphase}, and in all subsequent figures illustrating the dynamics
of Q-balls,  we plot the charge density (the integrand in equation
(\ref{q})) for the initial conditions and at later times\footnote{We
could have made similar plots of the energy density which are not equivalent.
We believe the charge density gives a  better
representation of the dynamics.}. The
parameter values used for this simulation are $\omega=1.5$ and $a=4$,
with the Q-balls initially in phase so that $\alpha=0.$ The two Q-balls
slowly attract and coalesce to  form one larger Q-ball which has a charge
which  is slightly less than the sum of the charges of the two
original Q-balls; the charge deficit being  carried away by the fission
of two additional Q-balls which, in figure~\ref{inphase}, can just be
seen moving away from the almost stationary large Q-ball at the origin. The attraction of the two Q-balls is simple to explain;
it being a consequence of the ratio $E/Q$ decreasing as $Q$
increases. However, the process of fission is less intuitive and is a
novel  concept to those who might have studied the dynamics of
topological solitons in an attractive potential. In the
topological  case the charge is an integer and so the fission of
higher charge solitons can only be achieved when the solitons are
moving sufficiently fast for the kinetic energy to overcome the
attraction and release a soliton. But here the charge of an isolated Q-ball can have any
value, arbitrarily close to zero, and so one might imagine that the
energy barrier to  fission at low interaction speeds is small, particularly when the charge is high.

\begin{figure}[ht]
\begin{center}
\leavevmode \epsfxsize=12cm\epsffile{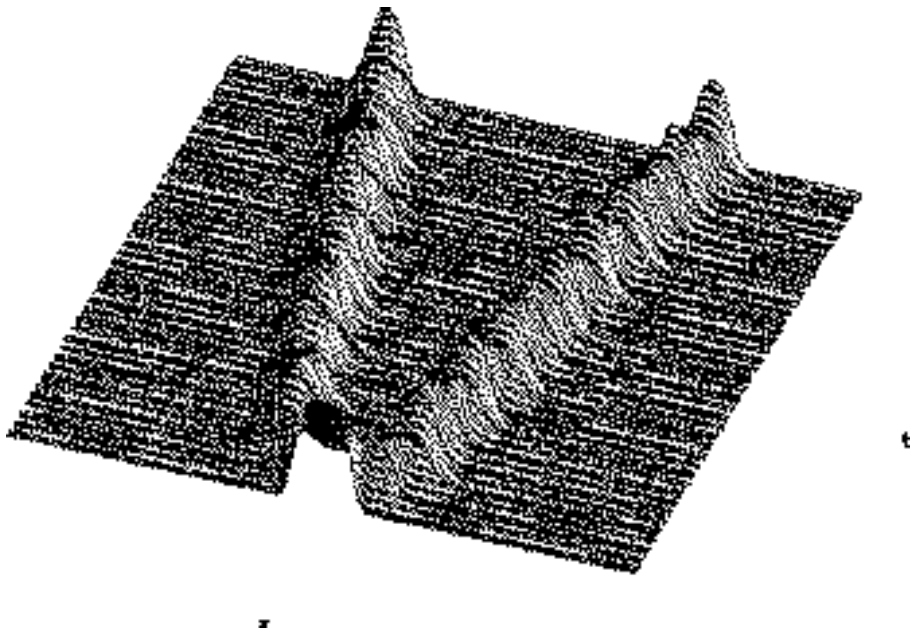}
\caption{The charge density at times $0\le t\le 100$, for a Q-ball with
$Q=8.4$ and a scale distortion $\lambda=1.6.$ Notice that the Q-ball
splits up into two equal parts. }
\label{largesquash}
\end{center}
\end{figure}
\begin{figure}[ht]
\begin{center}
\leavevmode \epsfxsize=12cm\epsffile{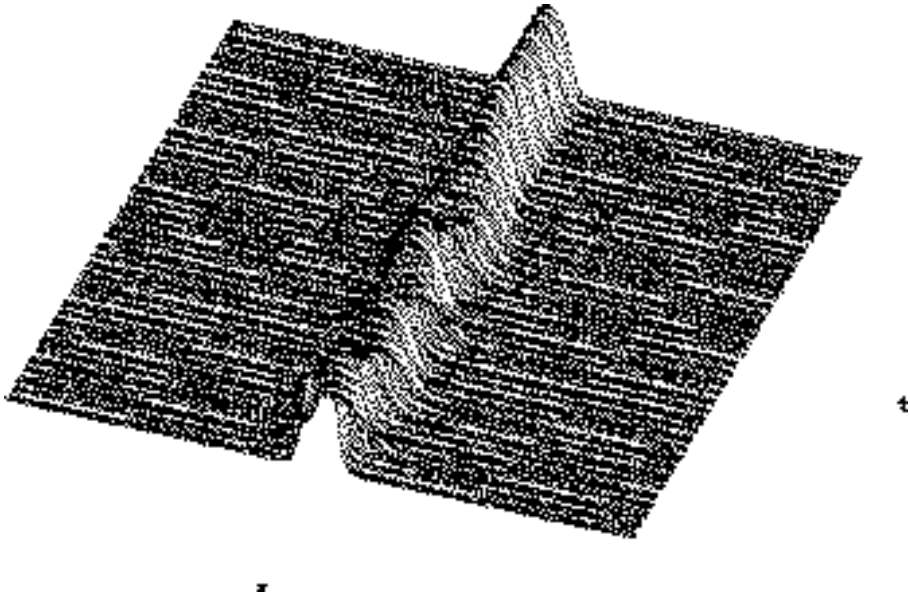}
\caption{The charge density at times $0\le t\le 3000$, for a Q-ball
with $Q=5.6$ and a scale distortion $\lambda=1.6.$ For this lower
charge the Q-ball does not fission and the oscillations damp with time.} 
\label{smallsquash}
\end{center}
\end{figure}

The fission of Q-balls is a process which can occur when a Q-ball suffers a
large distortion, for example, during a collision. This can be
demonstrated by taking a single Q-ball and squashing it by applying the
scale transformation 
\be x\mapsto \lambda x, \ \ \ \ \phi\mapsto\sqrt{\lambda}\phi\,,
\label{squash}
\ee 
where $\lambda>1$ is a scale factor. The scaling of the $\phi$
field is required in order that the charge of the Q-ball is not changed
by the scaling. For small enough values of $\lambda$ the Q-ball
oscillates, but then eventually returns to the original un-squashed Q-ball
corresponding to $\lambda=1$, which is to be expected since  these
Q-balls are stable. However, for a sufficiently large distortion the
Q-ball will break up into smaller Q-balls. This is illustrated in figure
\ref{largesquash}, where we take a Q-ball with charge $Q=8.4$ and
perturb it with a scale $\lambda=1.6.$

To consider the efficiency of the fission process as a function of the
charge, we define the quantity \be \Delta(Q)=(2E(Q/2)-E(Q))/E(Q) \,,\ee
where $E(Q)$ denotes the energy of a Q-ball with charge $Q.$
$\Delta(Q)$ is the fractional increase in energy required to allow a
charge $Q$ Q-ball to fission into two charge $Q/2$ Q-balls.  This is a
monotonically decreasing function of $Q$ with $\Delta(\infty)=0$,
with the limit $Q\rightarrow\infty$ being a Bogomolny-like limit in
which the energy is proportional to the charge. Thus we expect that
the fission of Q-balls  is more easily stimulated when the charge is
large.  To verify this we apply the same distortion factor
$\lambda=1.6$ as displayed in figure~\ref{largesquash}, but this time
we take a smaller Q-ball with charge $Q=5.6$,  and the results are displayed
in figure~\ref{smallsquash}. The Q-ball performs a breather-like motion
in which two structures initially begin to form but then recombine.
This motion persists for many cycles with a slowly decreasing
amplitude until it eventually settles down to a configuration which is
very close to the original Q-ball without a distortion (the
configuration at $t=3000$ is almost identical to the  stationary
$Q=5.6$ Q-ball). 

The above expectations of the fission of Q-balls are confirmed by
performing simulations, as in figure~\ref{inphase}, with two
stationary Q-balls and varying the charge (increasing or decreasing the
value of $\omega$). The charge of the additional Q-balls produced decreases
as the charge of the initial Q-balls is reduced  and for small enough
Q-balls no fission takes place; rather the sole Q-ball formed
oscillates for some time, with a decreasing amplitude.

\begin{figure}[ht]
\begin{center}
\leavevmode
\epsfxsize=12cm\epsffile{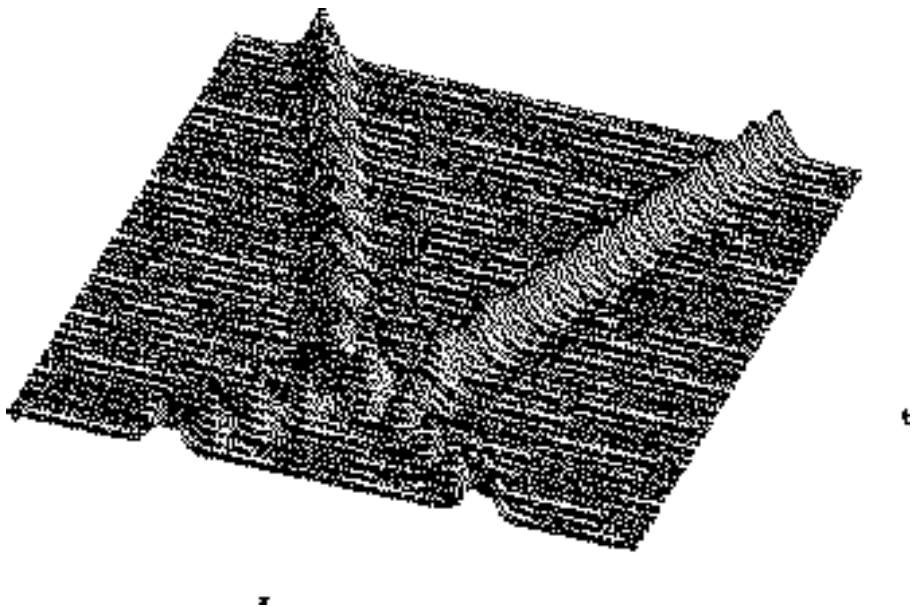}
\caption{The charge density at times $0\le t\le 150$,
for the parameters $\omega_1=\omega_2=1.5$, $a=10$, $\alpha=0$,
$v=0.3.$ The Q-balls pass through each other at this interaction speed and
coalesce at lower speeds.}
\label{boost}
\end{center}
\end{figure}

\begin{figure}[ht]
\begin{center}
\leavevmode
\epsfxsize=12cm\epsffile{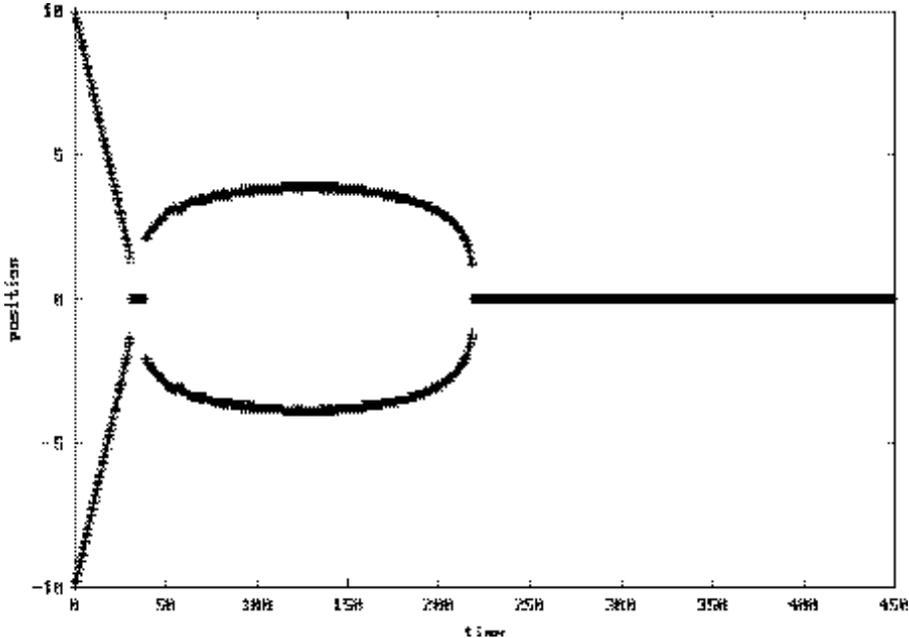}
\caption{Q-ball positions as a function of time for an
initial velocity of $v=0.28$ and all other parameters as in figure
\ref{boost}. We see that the Q-balls do not initially coalesce but have
insufficient energy to escape as in the case of $v=0.3$ and eventually
coalesce at the second attempt.} 
\label{position}
\end{center}
\end{figure}

If the two Q-balls are initially Lorentz boosted toward one another, each with a
velocity $v$ say, then if $v$ is large enough the two Q-balls can be
made to pass through each other. In figure~\ref{boost} we display the
charge density  for a simulation with $\omega_1=\omega_2=1.5$, $a=10$ and $v=0.3.$
In this case the two Q-balls pass through each other, although they do
lose some charge via radiation during the interaction process and
their velocities are reduced. For a slightly lower value
of the velocity the two Q-balls pass through each other,
but do not escape to infinite separation. Rather they subsequently 
recombine, forming a stationary Q-ball at the origin and producing two
additional Q-balls in the same manner as described above. In figure
\ref{position} we plot the positions of the two main Q-balls
(determined as the location of the maximum of the charge density) as a function of time for the velocity $v=0.28.$

\begin{figure}[ht]
\begin{center}
\leavevmode
\epsfxsize=12cm\epsffile{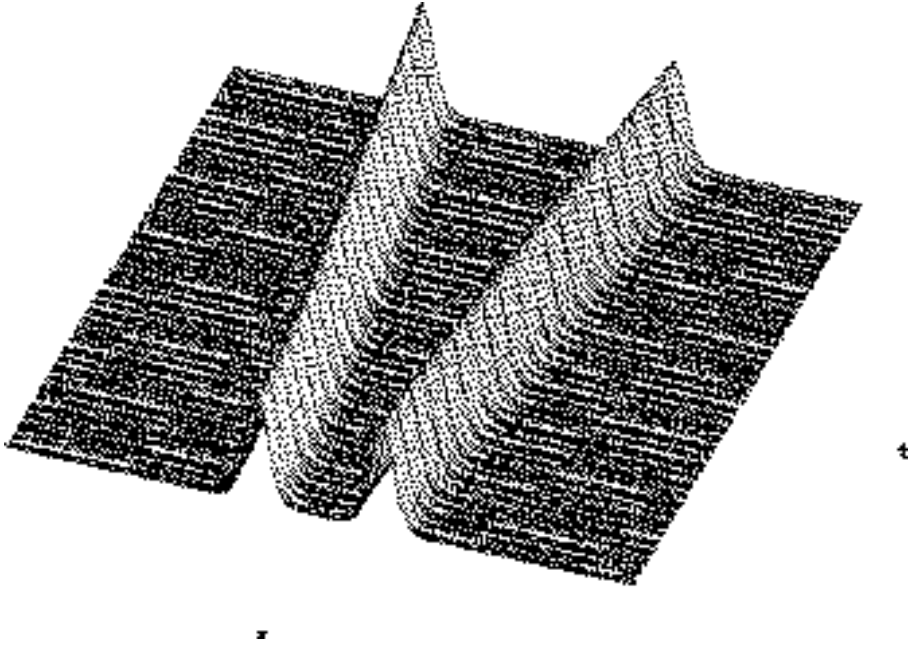}
\caption{As figure~\ref{inphase} except $\alpha=\pi.$ This is the
repulsive channel of Q-ball interactions.}
\label{outphase}
\end{center}
\end{figure}

To study how the relative phase affects the interaction we consider the
same initial configuration used to produce figure~\ref{inphase} except
that we set $\alpha=\pi$, so that the two Q-balls are exactly out of phase.
In this case the resulting evolution is very different and the two Q-balls
simply drift apart with no change in their shape or charge, even
though the crude arguments based on $E/Q$ suggest that they should
attract. The effects of changing the overall phase are similar in many
ways to the overall isospin rotations possible in 2-skyrmion
interactions. There it is possible to make the skyrmions attract or
repel by an internal SU(2) rotation about the line joining the
two soliton centres. Another way of understanding the relative phase
is as a current between two charged bubbles, if the Q-balls are thought of as
bubbles. The results of this repulsive interaction channel are 
displayed in figure~\ref{outphase}.

\begin{figure}[ht]
\begin{center}
\leavevmode \epsfxsize=12cm\epsffile{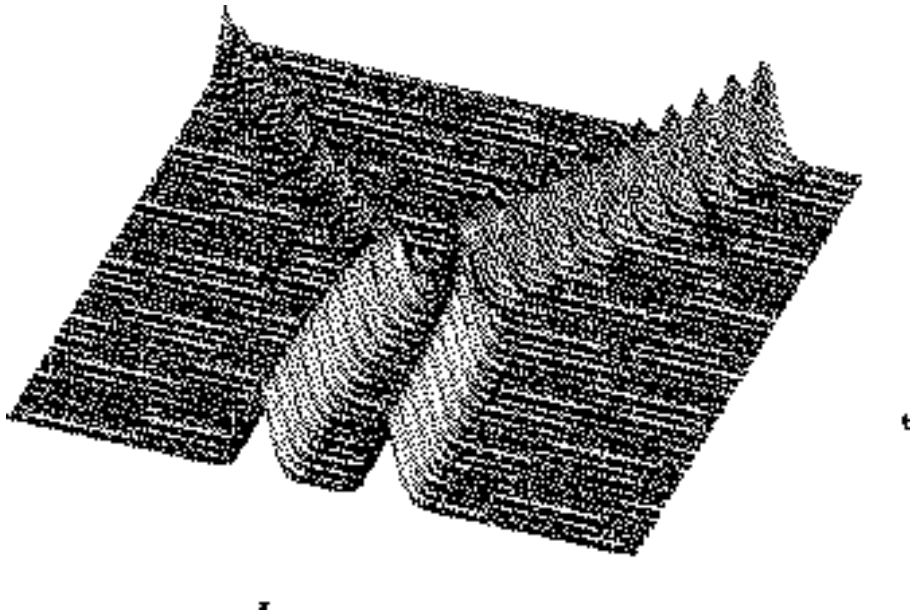}
\caption{As figure~\ref{inphase} except $\alpha=\pi/9.$ Notice the
novel process of charge transfer where the soliton in the left hand
half plane loses charge to the one in the right hand half
plane. Clearly the one which has lost charge is moving faster than the
other.} 
\label{piby4}
\end{center}
\end{figure}
\begin{figure}[ht]
\begin{center}
\leavevmode
\epsfxsize=12cm\epsffile{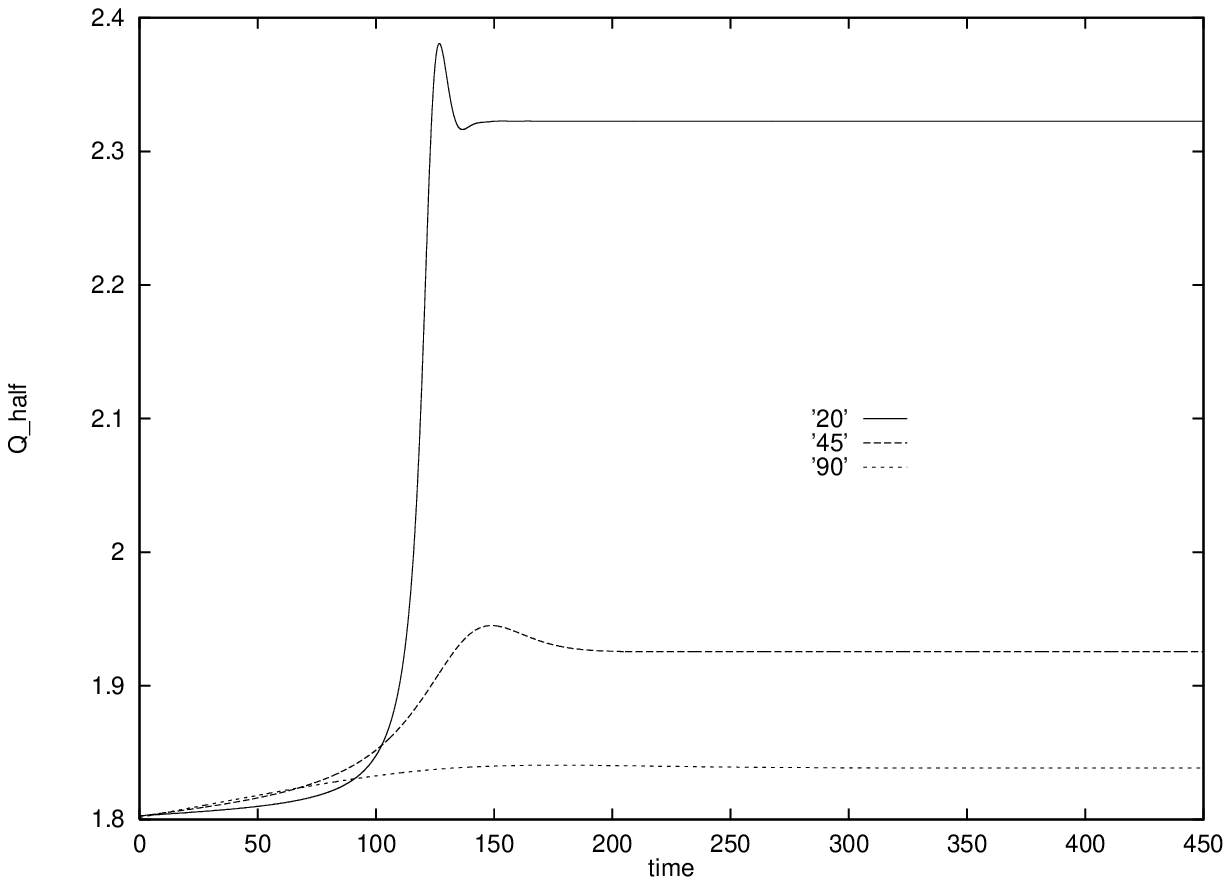}
\caption{The charge $Q_{\rm half}$ on the half-line $x>0$ as a function
of time for initial relative phases $\alpha=\pi/9,\pi/4,\pi/2.$}
\label{qhalf}
\end{center}
\end{figure}

As we have seen, for $\alpha=0$ the Q-balls attract and for
$\alpha=\pi$ they repel. However, for intermediate values of $\alpha$
the dynamics is much more complicated. We display the evolution for
$\alpha=\pi/9$ in figure~\ref{piby4}, illustrating a novel process
which we shall call charge transfer. The Q-balls initially move very
slightly towards each other, but they eventually repel and separate off
to infinity. The most interesting aspect is that the charge of the
first Q-ball has clearly decreased and that of the second Q-ball has
increased, despite the fact that the Q-balls remain two distinct
objects with only a very small overlap throughout the time evolution.
As they separate the smaller Q-ball moves at a greater speed than the
larger one.

In figure~\ref{qhalf} we plot the total charge $Q$ in the
 right half of the line, $x>0,$
as a function of time for the simulation displayed in figure~\ref{piby4},
 where $\alpha=\pi/9$,
and also for the cases $\alpha=\pi/4$ and $\alpha=\pi/2.$ We see that as the
relative phase is decreased the rate at which the charge transfer initially
takes place is reduced but the total charge transfered is increased. The
in-phase limit $\alpha=0,$ where the two Q-balls form a single larger Q-ball,
is a smooth limit if we interpret it as a total charge transfer. 
In the out-of-phase limit $\alpha=\pi$, as we have already seen,
there is no charge transfer.
If $\alpha<0$ then the same amount of charge is transferred as in the case
of a phase $-\alpha$, but this time it is the Q-ball in the left half of the
interval which increases in charge.

It may seem surprising that there is charge transfer, but this
result can be understood, at least qualitatively, by considering a
simplified mechanical analogue of the field dynamics associated
with two well-separated Q-balls.
Consider two equal charge Q-balls, fixed at the positions $x=\pm a$, 
then the ansatz (\ref{ansatz1}) may
be written in the form
\be
\phi=e^{i\theta_1}f(\vert x+a\vert)+e^{i\theta_2}f(\vert x-a\vert)\,,
\label{ansatz2}
\ee
where $\theta_1\equiv\theta_1(t),\theta_2\equiv\theta_2(t)$ are the time-dependent phases of the two Q-balls
and $f$ is a profile function.
To leading order in the separation $a$, corresponding to the
limit of large separation, the  contribution to the
Lagrangian is given by 
\be
{\cal L}=\frac{1}{2}M(\dot\theta_1^2+\dot\theta_2^2)
-\epsilon^2\cos(\theta_1-\theta_2)-4M\,,
\label{mech}
\ee
where 
$M=\int_{-\infty}^\infty f(\vert x\vert)^2\ dx$
 is treated as a constant moment of inertia and
$\epsilon^2= 4\int_{-\infty}^\infty f(\vert x+a\vert)f(\vert x-a\vert)\ dx$
is a small interaction coefficient. To derive this Lagrangian we have made
the assumption that the profile function is time independent, which obviously
has a very limited range of validity as we shall discuss further
below; it is nonetheless instructive.
The equations of motion which follow from (\ref{mech}) are
\be
\ddot\theta_1+\ddot\theta_2=0\,, \ \ \
\ddot\theta_1-\ddot\theta_2=\frac{2\epsilon^2}{M}\sin(\theta_1-\theta_2)\,.
\label{phasedynamics}
\ee
The first of these equations simply represents the fact that
the sum of the rotation frequencies $\dot\theta_1+\dot\theta_2$ is conserved, and
the second equation determines the dynamics of the relative phase.
There are symmetric solutions, $\theta_1=\theta_2$, and 
$\theta_1=\theta_2+ \pi$, where the phase difference remains constant,
corresponding to the two Q-balls being exactly
in-phase or exactly out-of-phase for all time, but for general values
of the initial relative phase, $\alpha=\theta_1(0)-\theta_2(0)$,
 there will be a non-trivial time dependence.
For all $\alpha\in(0,\pi)$ there will be an initial positive acceleration
in the relative phase, so that $\dot\theta_1>\dot\theta_2$ for 
small $t>0.$ Thus the first Q-ball will have a higher frequency than
the second Q-ball and, since we know that the charge of a Q-ball decreases with increasing frequency,
then this corresponds to the charge of the first Q-ball decreasing and the
charge of the second Q-ball increasing. This simple analysis also predicts that
the initial rate of charge transfer will be greatest for a relative phase
$\alpha=\pi/2$ and will decrease as $\alpha$ decreases.
This agrees with the observation we made earlier by an examination
of the plots in figure~\ref{qhalf} for small times. 

However, what we are clearly not able to study with our simple
restricted mechanical model is the  whole charge transfer process for
later times. One reason for this is that  we assumed that the profile
function $f$ was fixed when of course we know that it is highly
dependent on the rotation frequency (see equation (\ref{pro1d})). In
particular this dependence constrains the rotation frequencies to
satisfy $\omega_-<\dot\theta_1,\dot\theta_2<\omega_+$ and as either of
these limits are approached our simple model breaks down. One might be
tempted to improve our simple model to deal with this issue by
including the known frequency dependence of the profile function, but
it is not obvious how to do this since the profile function depends
upon the frequency $\dot\theta$ so using such an ansatz in the
Lagrangian would lead to a Lagrangian for a mechanical system with
second order derivatives $\ddot\theta$ and hence a fourth order
equation of motion. Furthermore, we have assumed that the positions of
the Q-balls are fixed when in fact the results of the full simulations
show that they eventually drift apart. This effect will also serve to
cut-off the relative phase dynamics since it will correspond to
reducing the $\epsilon^2$ coefficient in our simple mechanical model. 

In summary, we have shown that a simple  mechanical model is useful in
understanding the qualitative features of the charge transfer process,
but a more sophisticated analysis is required to explain the
quantitative behaviour found. The analysis of relative phase dynamics
in mechanical systems, such as discrete breathers,  has been studied in
some detail and the phase space trajectories are well understood
\cite{ams}. These methods can be extended to study
the more complicated relative phase dynamics, and hence charge transfer, 
of Q-balls~\cite{BMS}.

\begin{figure}[ht]
\begin{center}
\leavevmode
\epsfxsize=12cm\epsffile{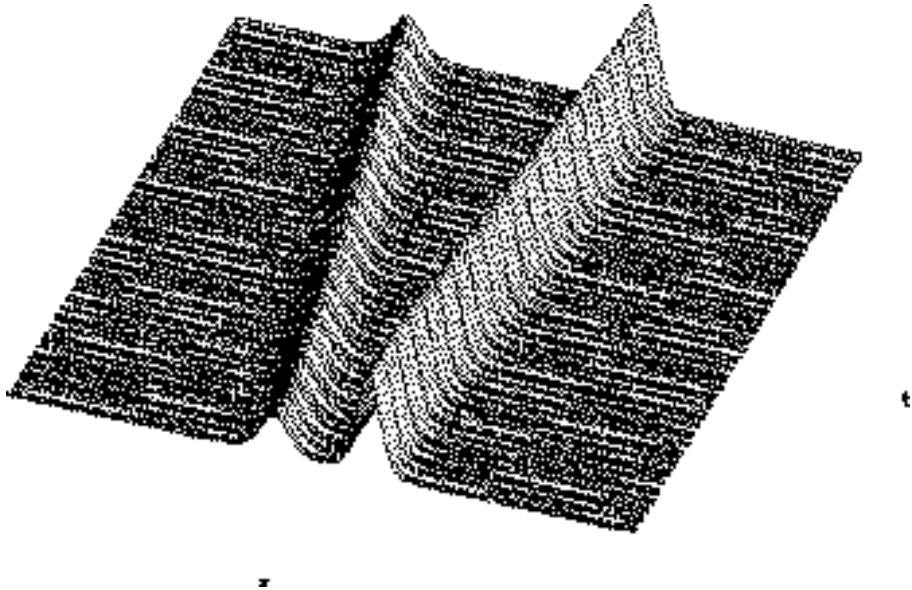}
\caption{The charge density at times $0\le t\le 450$ for the parameters
$\omega_1=1.8, \omega_2=1.5, a=3, \alpha=0.$ The two Q-balls repel with
virtually no charge transfer since they never get close enough.} 
\label{unequal}
\end{center}
\end{figure}
   \begin{figure}[ht]
\begin{center}
\leavevmode
\epsfxsize=12cm\epsffile{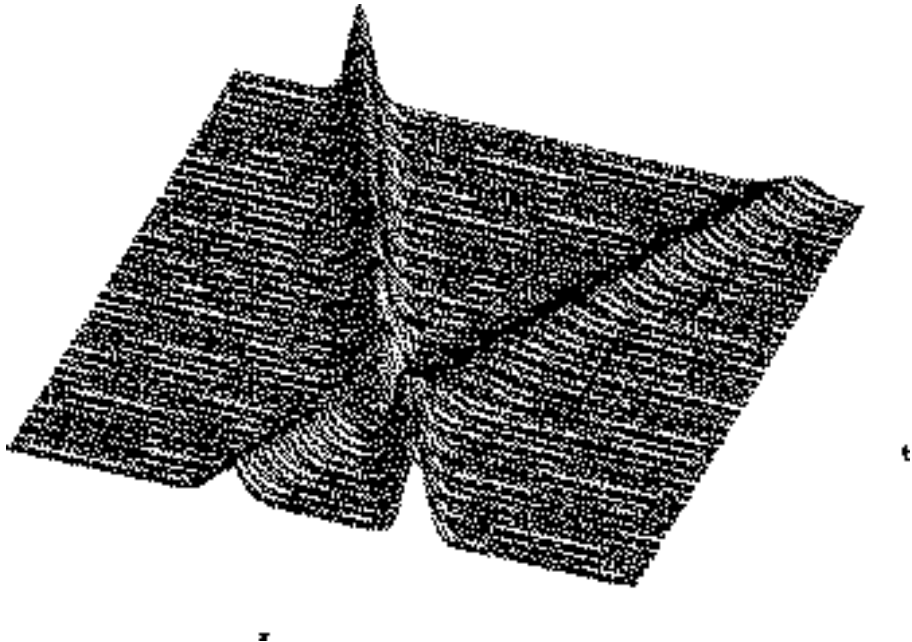}
\caption{The charge density at times $0\le t\le 95$ for the 
parameter values $\omega_1=1.8, \omega_2=1.5, a=6, \alpha=0, v=0.2.$
The non-zero relative velocity allows the interaction and charge transfer
takes place.}
\label{unequalboost}
\end{center}
\end{figure}

So far we have only considered initial conditions in which the two
Q-balls have the same charge. For two Q-balls which have different charges
 the initial relative phase does not have the same importance as for
Q-balls of the same charge, due to the fact that the Q-balls have
different frequencies and so the initial relative phase will not be
preserved, even with no interaction. This can easily been seen using
the simple mechanical system above. 

In figure~\ref{unequal} we plot the charge density for the initial
conditions  $\omega_1=1.8, \omega_2=1.5, a=3, \alpha=0.$ It can be
seen that the two Q-balls repel and there is virtually  no charge
transfer since the solitons never get close enough for the charge
transfer process to become important. Similarly, if a non-zero
relative phase is introduced then virtually no charge transfer takes
place, although the rate of separation does vary very
slightly. However, if the Q-balls are Lorentz boosted towards each
other with a sufficiently large velocity $v$ so that they collide and
pass through each other, it is possible to induce charge transfer as
illustrated in figure~\ref{unequalboost} for the parameters
$\omega_1=1.8, \omega_2=1.5, a=6, \alpha=0, v=0.2.$  The amount of
charge transferred depends on the value of the relative phase as the
Q-balls collide, as can be verified by changing the initial phase.  It
can be checked that this is equivalent to varying the initial
separation, since the time to collision is then altered and hence the
relative phase is different by an amount equal to the change in
collision time multiplied by the frequency difference. Just using the
simple mechanical analogy, one might have naively expected that an
initial difference in the rotation speeds  would be on
 a similar footing to an
initial phase difference, but this is clearly not the case. It is
evident that there is a non-trivial interaction between the relative
dynamics of the Q-balls and the charge transfer process.

\section{Planar Q-balls}\news\ \ \ \ \ \

The main features of one-dimensional Q-balls which we have described in
the previous section, such as charge transfer and the dependence of
the interaction force on the relative phase, carry through to the
two-dimensional case. We demonstrate this by again performing
numerical simulations of the field equations via an equivalent finite
difference scheme to the one dimensional case. We find that a grid
containing $200^2$ points with $\Delta x=0.2$ and $\Delta t=0.05$  gives
an accurate representation of the dynamics in this case.  In
contrast to the one-dimensional case  an exact solution is not known
for the profile function in two-dimensions, but it is a simple matter
to numerically obtain the profile function using a standard shooting
method.  

\begin{figure}
\begin{center}
\leavevmode
\epsfxsize=7cm\epsffile{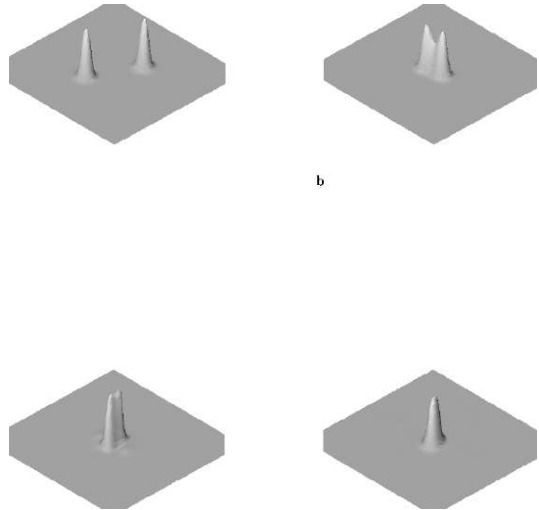}
 \caption{The charge density at $t=0,104,112,3200$ for two Q-balls with
$\omega_1=\omega_2=1.6$, positions $\pm(6,3)$, velocities $v=0.05$, and
relative phase $\alpha=0.$ This interaction with non-zero impact parameter 
shows the attractive potential of the two Q-balls, which coalesce into a
larger Q-ball.}
\label{d10}
\end{center}
\end{figure}

One might assume that head-on collisions of Q-balls with a small charge
(for example, $\omega=1.6$) are equivalent to those in  one-dimension
with attraction, repulsion and charge transfer taking place  as before
and indeed this is the case as we will discuss later.  The
phase-dependent force is, however, a more general concept which we  will
illustrate in fully two dimensional interactions with a non-zero impact
parameter.  In particular, we consider the collision of two equal
charge Q-balls ($\omega_1=\omega_2=1.6$) with a non-zero impact
parameter, where the initial positions of the Q-balls are
$(x_1,x_2)=\pm(6,3)$, each is boosted along the $x_1$-axis with a
velocity $v=0.05$, and the relative phase $\alpha$  is set to zero. In
figure~\ref{d10} we plot the charge density for this interaction  at
$t=0,104,112,3200.$ 

\begin{figure}
\begin{center}
\leavevmode
\epsfxsize=7cm
\epsffile{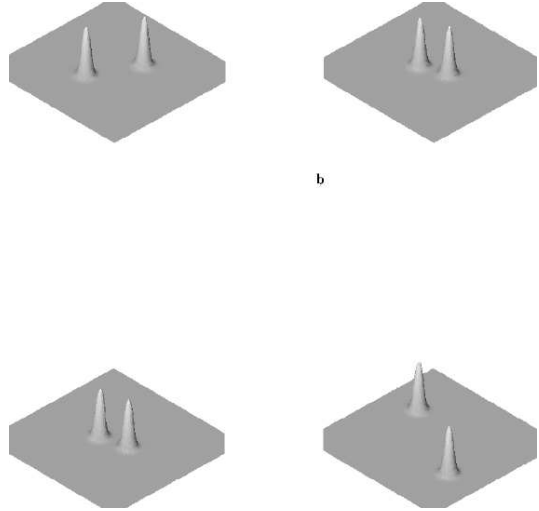}
\ \vskip 0cm \caption{The charge density at $t=0,104,140,200$ for all
parameters as 
in figure~\ref{d10} except $\alpha=\pi.$ We now see the repulsive interaction
of the two Q-balls as in the one-dimensional case, where it could only be observed through head-on collisions.}
\label{d11}
\end{center}
\end{figure}

\begin{figure}
\begin{center}
\leavevmode
\epsfxsize=7cm\epsffile{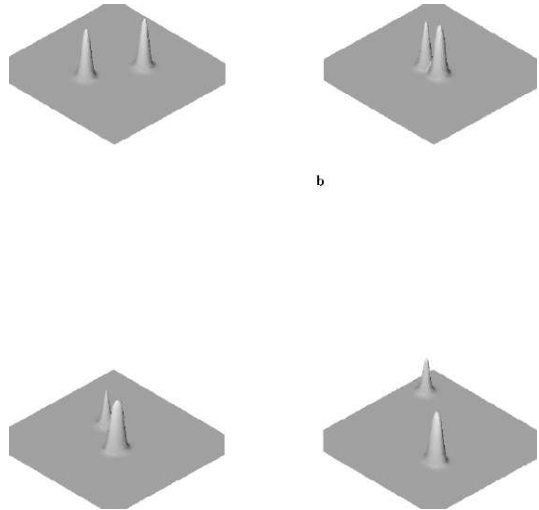}
 \caption{As figure~\ref{d11} except $\alpha=\pi/4.$ Charge 
transfer takes place in an almost analogous way to the one-dimensional 
interactions.}
\label{d09}
\end{center}
\end{figure}

\begin{figure}
\begin{center}
\leavevmode
\ \vskip -3cm
\epsfxsize=12cm
\centerline{\epsffile{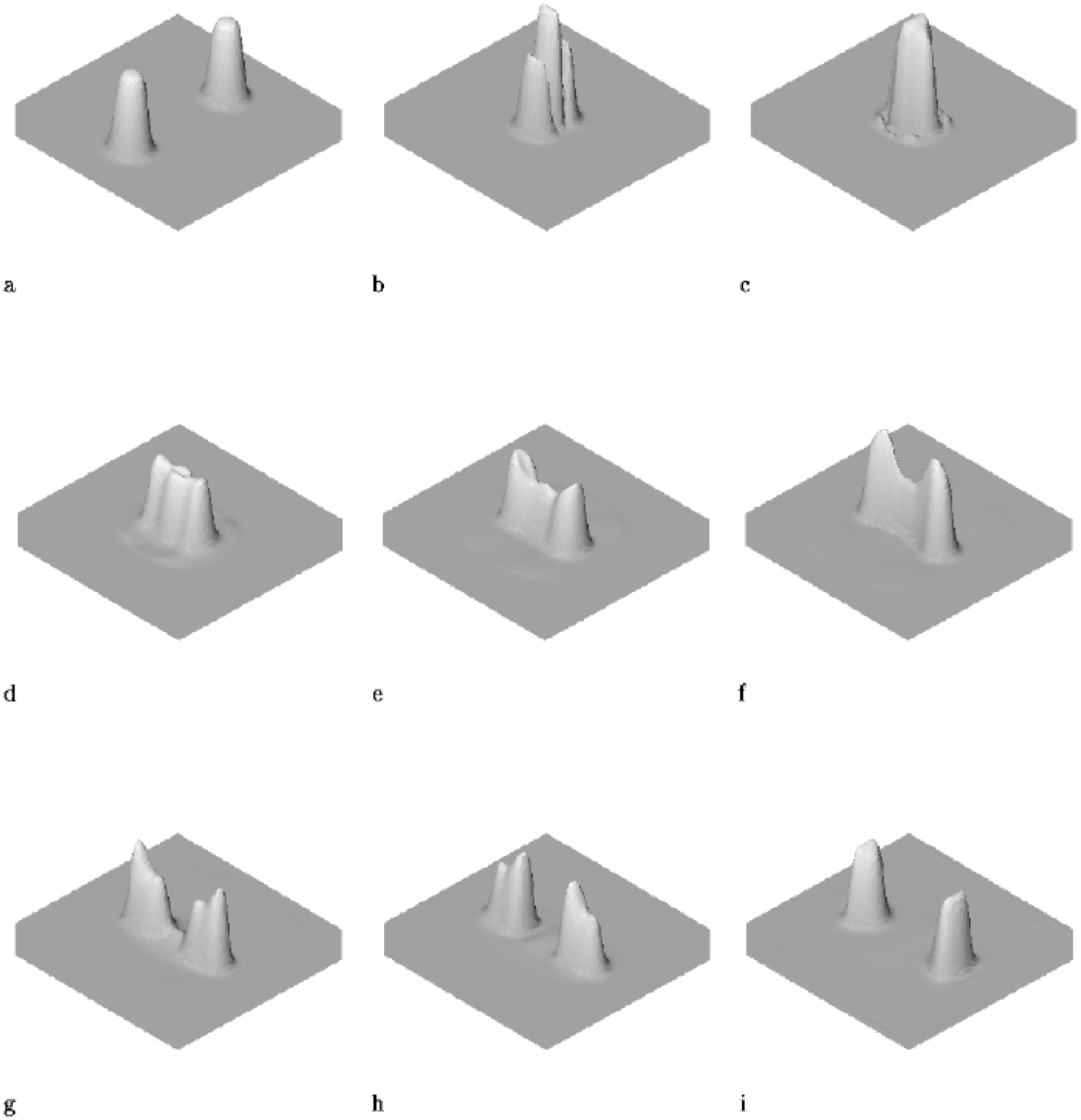}}
\caption{The charge density at $t=0,20,24,28,32,36,40,44,52$ for two Q-balls
with $\omega_1=\omega_2=1.5$, positions $\pm(10,0)$, velocities $v=0.4$ and
relative phase $\alpha=0.$ The two Q-ball under go a complicated interaction 
process which eventually leads to them being scattered at right angles to the 
incident direction.}
\label{d05}
\end{center}
\end{figure}

As in the one-dimensional case (where collisions are head-on) the two Q-balls
attract and form a single larger Q-ball, although the large Q-ball has some 
angular momentum due to the fact that the collision was not head-on.
Figure~\ref{d11} displays the results of the same simulation except
that the two Q-balls are exactly out-of-phase, that is $\alpha=\pi,$
where the Q-balls clearly repel. Finally, in figure~\ref{d09}
we display the simulation with a relative phase $\alpha=\pi/4$, where
there is an initial attraction, followed charge transfer and 
finished off by a repulsion which forces the two Q-balls apart. We conclude,
therefore, that while the head-on collisions of small Q-balls in two-dimensions
can be thought of as being effectively one-dimensional, the same 
dynamical processes are active in the case of a non-zero impact parameter.

Next, we turn our attention to head-on collisions of Q-balls with
higher charge,  where --- based on the intuition of the
one-dimensional interactions --- one would expect things to be
slightly different.  We take two Q-balls with $\omega_1=\omega_2=1.5$
at positions $(x_1,x_2)=\pm(10,0)$ with each Lorentz boosted towards
the other with a velocity $v=0.4.$ Figure~\ref{d05} displays the
charge density at $t=0,20,24,28,32,36,40,44,52$ for the in-phase case,
$\alpha=0.$ As can be seen from the figure, there is a very
complicated interaction process involving the charge being strongly
deformed and the emission of some radiation. Eventually, two Q-balls
emerge from the  interaction region at right angles to the initial
direction of approach.  Naively one may think that this is a simple
$90^\circ$ scattering process as seen in a number of topological
soliton  models~\cite{vort,mon,sky} and suggested for Q-balls  in
ref.~\cite{greek}\footnote{We should note that the work presented in
ref.~\cite{greek} uses a potential of type II, not type I as used in
our work, but that the qualitative nature of
this process is independent of
the choice of potential.}. However, the scattering of Q-balls is a  complicated
dynamical issue rather than being topological, and the underlying
mechanisms are very different. In particular, there is no associated
geometry of a moduli space which forces the Q-balls to scatter at right
angles.  Rather, during collisions the Q-matter becomes highly
deformed with huge charge densities and it is this deformation, and
its associated pressures, that lies at the heart of the interaction
process and the fission of Q-balls in the plane perpendicular to the
incident direction.  As we demonstrated in the one-dimensional case, a
sufficient distortion of a Q-ball will induce fission, and it this same
process which is responsible for this more complicated phenomena in
two-dimensions.

\begin{figure}
\begin{center}
\leavevmode
\ \vskip -3cm
\epsfxsize=12cm
\centerline{\epsffile{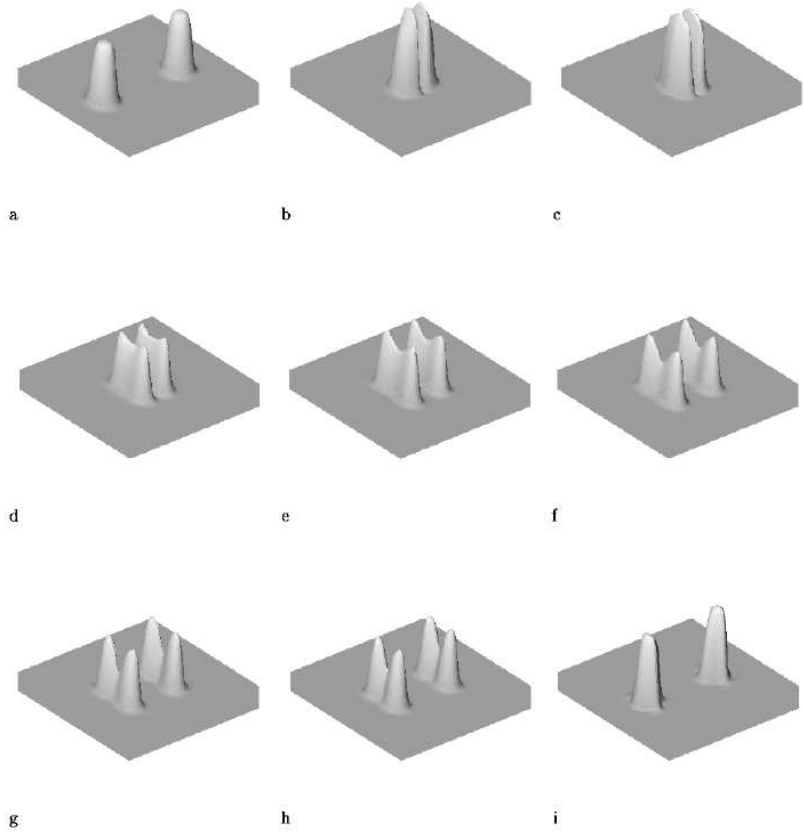}}
\caption{As figure~\ref{d05} except $\alpha=\pi.$ The two Q-balls come together,
but never coalesce. Due to their large charge,
 distortion of the Q-matter takes place
inducing the fission of four Q-balls in the plane perpendicular to the incident
direction.}
\label{d13}
\end{center}
\end{figure}

This point can be illustrated immediately by considering the same
scattering process, but this time we set the Q-balls to be exactly
out-of-phase, that is $\alpha=\pi$ with the results displayed in 
figure~\ref{d13}. Although the initial conditions look identical
in figures \ref{d05} and \ref{d13}, the evolution is clearly very
different. As the two Q-balls are now in a repulsive phase
the Q-matter gets distorted in a very different way. Rather
than forming a single structure, as in figure~\ref{d05}c, the two
individual Q-balls never actually coalesce because of the repulsive 
interaction, getting squashed separately and this distortion
induces the fission of each. Thus, each Q-ball splits into two and the
two pairs repel each other, producing four Q-balls in all, which are
clearly visible in figure~\ref{d13}g. For this particular set of 
initial parameters the Q-ball pairs do not have enough energy to escape
each others attraction and eventually recombine leaving
two Q-balls which move off to infinity. By, for example, increasing the
initial velocity it is found that the four Q-balls can be produced in such
a way that they all separately move off to infinity without
any subsequent recombination. If the initial velocity is small enough
then the distortion is not large enough to induce fission and the
two Q-balls eventually repel keeping their individual structure intact.

\begin{figure}
\begin{center}
\leavevmode
\ \vskip -3cm
\epsfxsize=12cm
\centerline{\epsffile{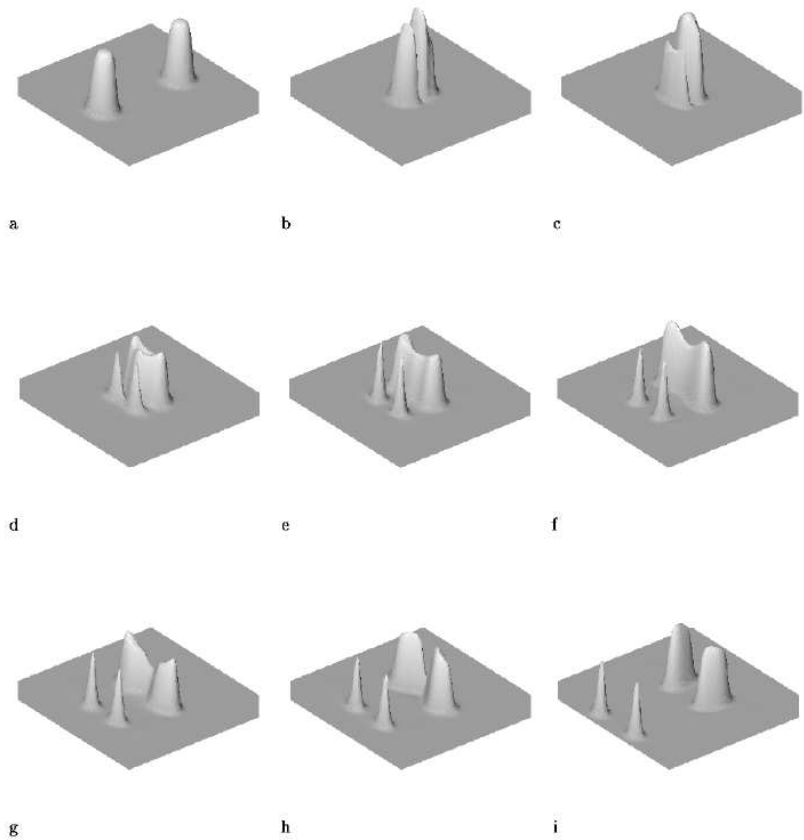}}
\caption{As figure~\ref{d05} except $\alpha=\pi/2.$ Now charge transfer takes
place, as well as fission into the plane perpendicular to the incident
direction.}
\label{d14}
\end{center}
\end{figure}

\begin{figure}
\begin{center}
\leavevmode
\ \vskip -3cm
\epsfxsize=12cm
\centerline{\epsffile{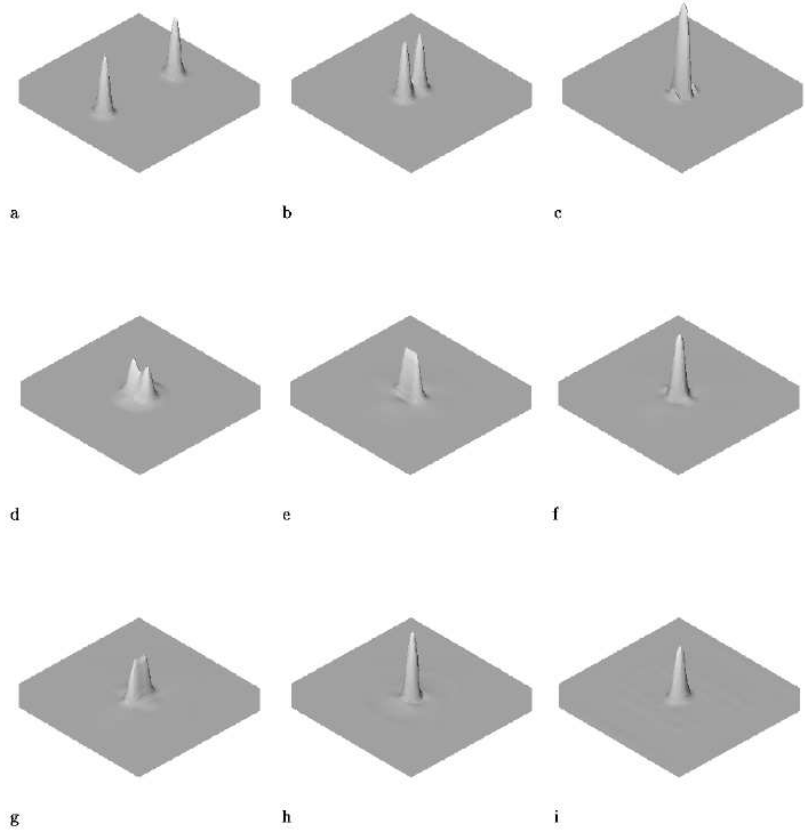}}
\caption{The charge density at $t=0,20,24,28,36,40,44,52,80$ for two Q-balls
with $\omega_1=\omega_2=1.6$, positions $\pm(10,0)$, velocities $v=0.4$ and
relative phase $\alpha=0.$ After some oscillations around the centre, the 
final configuration settles down to a single Q-ball at the centre.}
\label{d06}
\end{center}
\end{figure}

\begin{figure}
\begin{center}
\leavevmode
\epsfxsize=7cm
\centerline{\epsffile{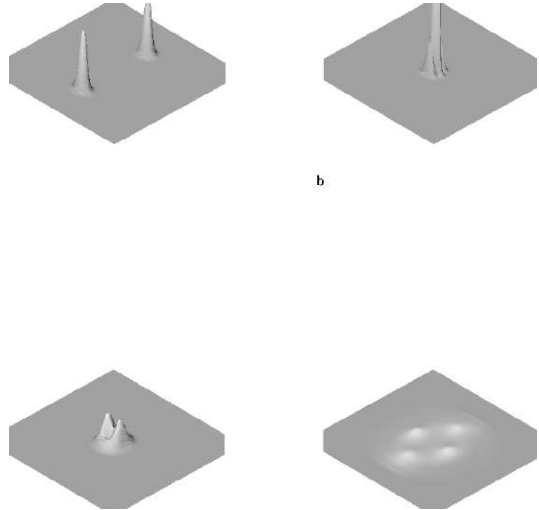}}
 \caption{The charge density at $t=0,16,20,32$ for two Q-balls
with $\omega_1=\omega_2=1.6$, positions $\pm(10,0)$, velocities $v=0.6$ and
relative phase $\alpha=0.$ The extra kinetic energy results in a highly 
inelastic collision, with four small Q-balls left at the end, plus a large
amount of radiation.}
\label{d07}
\end{center}
\end{figure}

\begin{figure}
\begin{center}
\leavevmode
\epsfxsize=7cm
\centerline{\epsffile{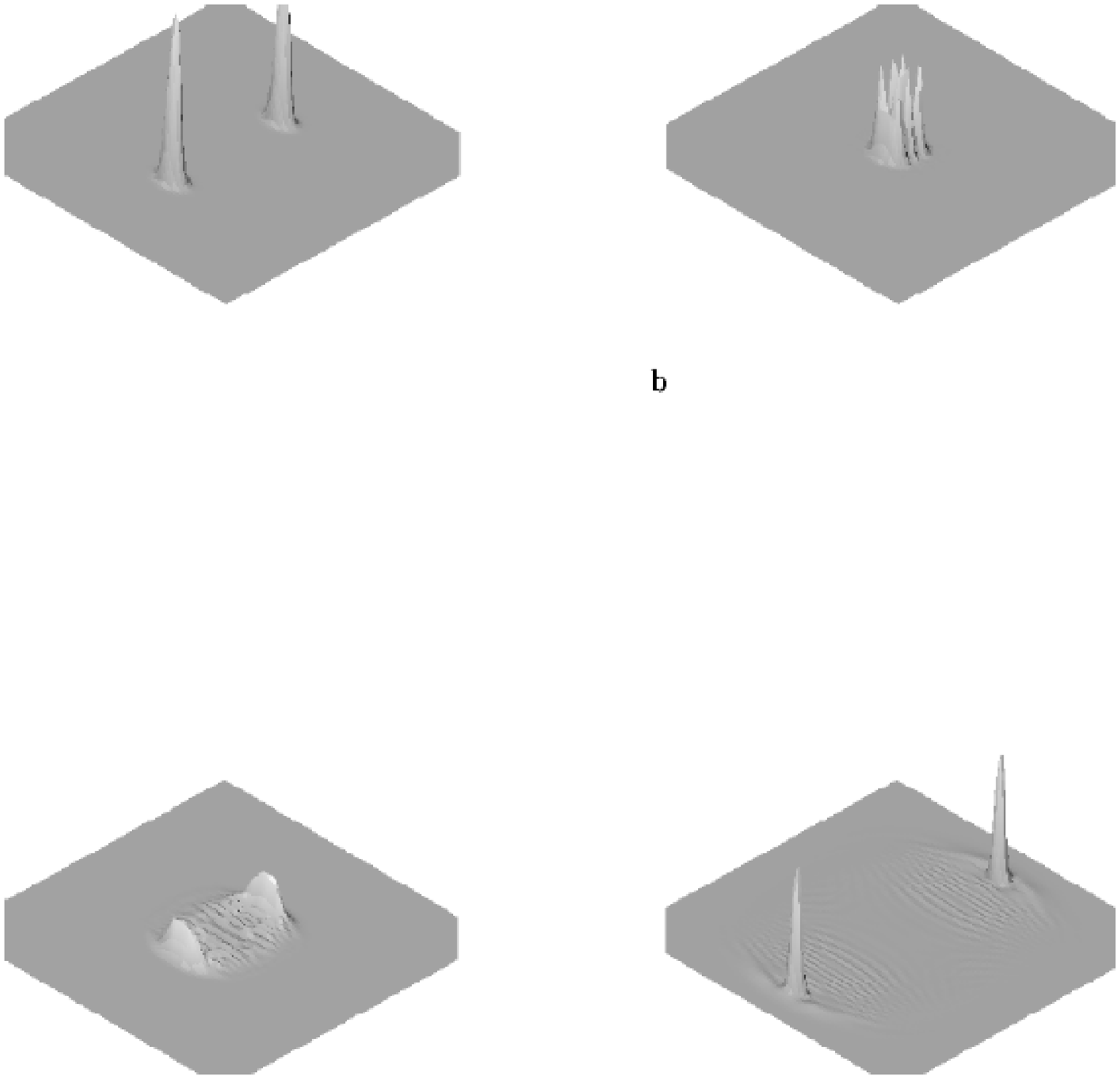}}
\caption{The charge density at $t=0,16,24,40$ for two Q-balls
with $\omega_1=\omega_2=1.6$, positions $\pm(10,0)$, velocities $v=0.8$ and
relative phase $\alpha=0.$ The Q-balls are moving so fast now that their 
momentum carries them through each other before they have time to interact.}
\label{d08}
\end{center}
\end{figure}

In figure~\ref{d14} we investigate the same simulation as in figure~\ref{d05} 
but with an
initial relative phase $\alpha=\pi/2.$ In this case we expect that the
distortion will also be accompanied by a charge transfer, and 
indeed this is what we find, as the first Q-ball loses charge to the
second one and then each Q-ball undergoes fission to produce a total of two
small Q-balls and two large Q-balls. The two small Q-balls move away from
the interaction region at a faster rate than the large Q-balls, and they
do not recombine. The two large Q-balls move away from the interaction
region at only a very slow speed and in fact they eventually recombine after
a time of around $t=200$ (the final plot in figure~\ref{d14} is only
at $t=52$).

As we discussed in the one-dimensional case, large Q-balls are more
susceptible to fission than smaller Q-balls, so we expect that the
scattering processes we have described above will vary depending on
the charge of the initial Q-balls. For example, our reasoning predicts
that the fission process in figure~\ref{d05}, which produced two
Q-balls moving at  right-angles to the initial line of approach, will
be more difficult to reproduce for smaller charges. To test this we
perform the same simulation, with $v=0.4$ again, but decrease the
charge of the initial Q-balls by taking $\omega=1.6$ rather than
$\omega=1.5.$ The resulting evolution is shown in figure~\ref{d06} and
confirms that now the distortion is not sufficient to liberate two
Q-balls. The configuration oscillates for some time before settling
down to a single larger Q-ball, after a small amount of charge has been
dissipated through radiation. Fission can be produced for these
smaller Q-balls by increasing the impact velocity, but this also has
the result that some of the charge passes straight through the
interaction region producing small Q-balls which continue to travel
along the direction of approach. A collision at increased velocities
is also a more violent process and more charge is lost to radiation in
these circumstances. In figure~\ref{d07} we display the evolution for
the case where the velocity is increased to $v=0.6$, with all other
parameters kept the same as in figure \ref{d06}. Just visible in
figure~\ref{d07}d  are the four very small Q-balls which are produced
by this collision together with a ring of charge carried away by the
radiation generated. If the collision velocity is increased further
then the two Q-balls have less time to interact and their momentum
carries them through the collision process with no deflection. This is
demonstrated in figure~\ref{d08} where $v=0.8$ and no additional
Q-balls are produced. In summary, as the charge is reduced an increased
velocity is required in order for a sufficient deformation to be
generated to produce fission, but this also results in  more of the
charge  being carried straight through the collision. Thus for small
charge there is a very limited window of velocities for which
collisions of the form displayed in figure \ref{d05} may occur.


\section{3D Q-balls}\news\ \ \ \ \ \

In the previous two sections we have built up a picture of the dynamics of
Q-balls in one and two dimensions. In going from the extensive study
in one dimension to two dimensions we have noted a number of subtle
effects associated with the extra dimension. However, the basic processes
involved are the same: attraction, repulsion and charge transfer. In
this section we will apply the same numerical techniques to the case
of three dimensions.  To begin with we conducted an extensive study of
the dynamics on grids containing $100^3$ points  and have once again
found that in many cases the dynamics are very similar to those in
one-dimension. At the risk of labouring  the point we found that for
small Q-balls, if they were initially in-phase they coalesced, while if
they were out-of-phase they repelled, and if they had any other phase
they engaged in  charge transfer. However, we did find some extremely
complicated interactions which are related  to those seen in the case
of two dimensions.  As was pointed out in the previous section when
the Q-balls have a large charge, their interactions can have some
interesting variants in two dimensions and it is these particular
cases in three dimensions on which we will focus in this section.

\begin{figure}
\begin{center}
\leavevmode
\ \vskip -3cm \hskip 0.5cm
\epsfxsize=12cm\epsffile{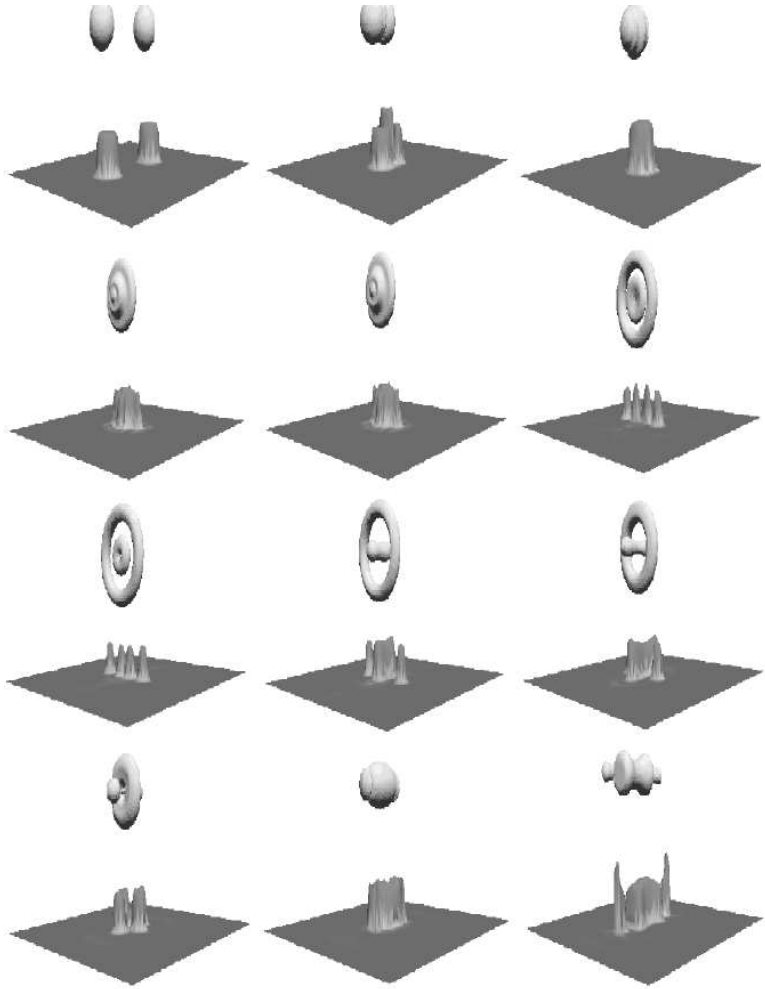}
\caption{The charge density at t=0,27,33,39,45,60,75,90,105,120,135,150
for two Q-balls with $\omega_1=\omega_2=1.5$, initial positions $\pm(15,0,0)$,
velocities $v=0.4$ and relative phase $\alpha=0$. Shown is the
three-dimensional isosurface and a two dimensional slice through the
centre of the interaction region which should be compared to the
corresponding two dimensional interaction in figure~\ref{d05}. Note
the production of a loop in the plane perpendicular to the line of
incidence, which expands before recollapsing into Q-balls along the
incident direction.}
\label{3d1}
\end{center}
\end{figure}

\begin{figure}
\begin{center}
\leavevmode
\ \vskip -3cm \hskip 0.5cm
\epsfxsize=12cm\epsffile{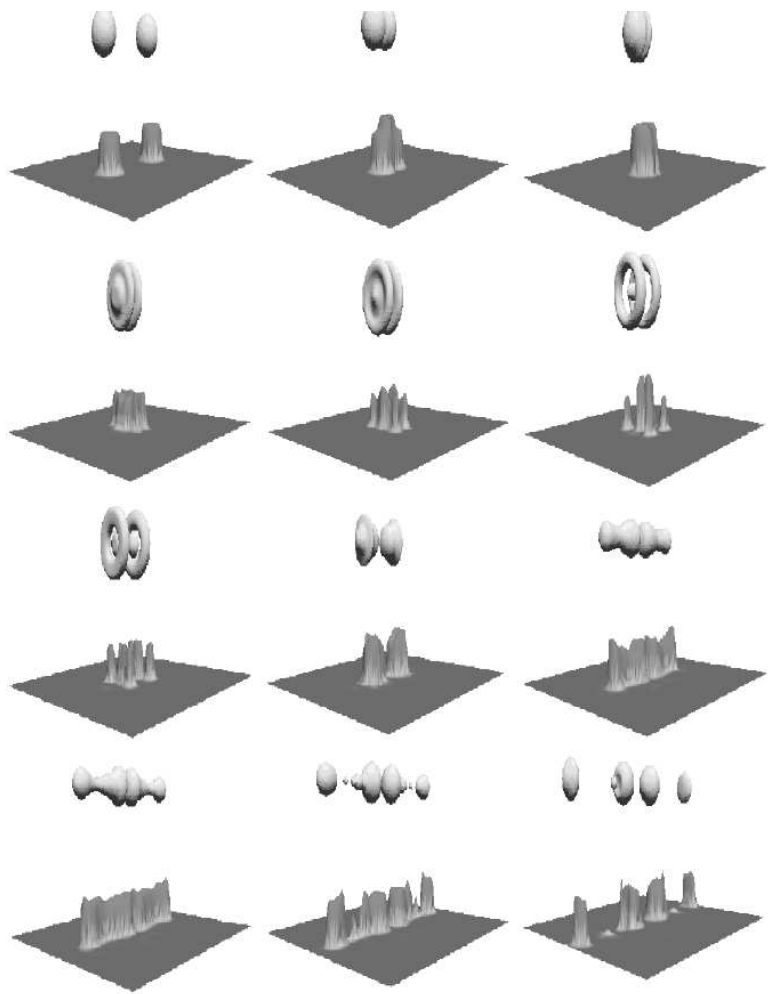}
\caption{As figure~\ref{3d1} except that $\alpha=\pi$, with the
equivalent two dimensional interaction being figure~\ref{d13}. Fission
of the Q-balls produces two loops in the plane perpendicular to the
incident direction, which are subsequently repelled while recollapsing
and emitting Q-balls along the incident direction.}
\label{3d2}
\end{center}
\end{figure}

\begin{figure}
\begin{center}
\leavevmode \ \vskip -3cm \hskip 0.5cm
\epsfxsize=12cm\epsffile{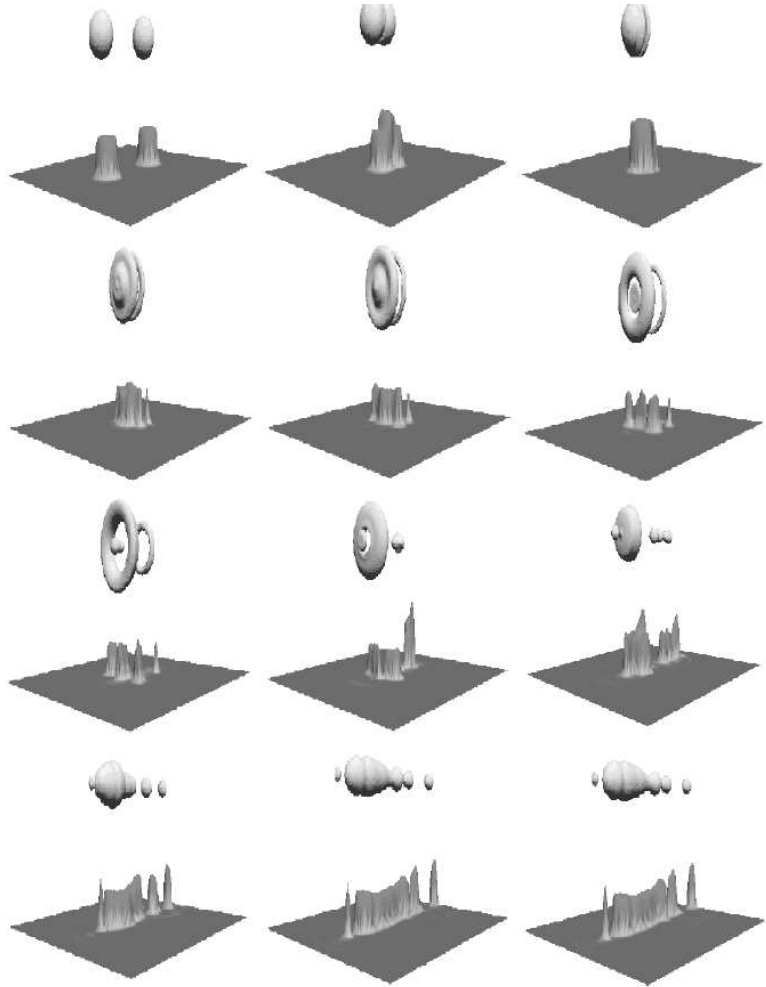}
\caption{As figure~\ref{3d1} except that $\alpha=\pi/2$, with the
equivalent two dimensional interaction being figure~\ref{d14}. Charge
transfer takes place during the interaction leading to the formation
of two loops which are not of equal charge. These two loops  are then
repelled, the smaller one having a higher speed.}
\label{3d3}
\end{center}
\end{figure}

In particular we have focussed on the analogues of figures~\ref{d05},
\ref{d13} and \ref{d14}  in which complicated dynamical phenomena were
identified. In each of these cases we have  performed the analogous
simulations on a grid of $300^3$ points with $\Delta x=0.3$ and
$\Delta t=0.03$ in order to accurately simulate the complicated
dynamical processes. In each of the three cases we start with 2 Q-balls
each with $\omega=1.5$ at $\pm(15,0,0)$, Lorentz boosted toward each
other with a velocity $v=0.4$. The results of the simulations are
displayed in figures~\ref{3d1}, \ref{3d2} and \ref{3d3}.

The in-phase case (figure~\ref{3d1}) has some marked similarities
to the equivalent case in two dimensions (figure~\ref{d05}) if one
looks at the two-dimensional slice through the centre of the
Q-balls\footnote{It should be noted that the two interactions are
not equivalent  even when the parameters are almost identical since
the relationship between the charge $Q$ and the frequency $\omega$ is
not the same in two and three dimensions.}. However, the extra dimension has
one remarkable effect: it allows for the production of a loop in the plane
perpendicular to line joining the two incident Q-balls. This phenomena
is the three dimensional analogue of the right-angled fission process
described in the two dimensional case. But in three dimensions the
fission takes place symmetrically in all directions in the plane while
respecting the cylindrical symmetry of the initial configuration. The
loop expands leaving some  charge in the centre which later expands to
create a second, much smaller loop. Later both the loops collapse and
Q-balls emerge back along the incident direction.

The formation of the loop in this three dimensional simulation adds
further weight  to our earlier discussion of the two dimensional case
where we pointed out that right-angled fission of two Q-balls was not
of topological origin, nor was it even related to the topological
interactions of, for example, vortices~\cite{vort},
monopoles~\cite{mon} and skyrmions~\cite{sky}. The reason being that
in a topological interaction in three dimensions one would have
expected there to have been a preferred direction (this is because in
the case of topological solitons, for example skyrmions, the field
configuration of a single soliton is not spherically symmetric,
although the change in the  field due to a spatial rotation can be
undone by acting with a symmetry of the theory, which means quantities
such as the energy density are spherically symmetric). But as we have
already pointed out the formation of a loop is reliant on all
directions being on an equal footing as far as the fission process is
concerned, which of course is due to the fact that the field itself is
spherically symmetric for a single Q-ball.

This explanation of the formation of a loop during the interaction of
two large Q-balls in three dimensions is compatible with the results
of the out-of-phase case (figure~\ref{3d2}) and that of a general
relative phase (figure~\ref{3d3}). In both cases, the two dimensional
slice through the interaction region is very similar to that of the
equivalent two dimensional interactions (figures~\ref{d13} and
\ref{d14} respectively). In the out-of-phase interaction, two
identical loops form which are then repelled back along the direction
from which they came. As they move away they begin to collapse, the
final outcome being a series of symmetrically placed Q-balls along the
incident direction. The interaction for $\alpha=\pi/2$ is similar,
except that, as expected, charge transfer takes place during the
interaction and the two loops created have very different charge.

\begin{figure}
\begin{center}
\leavevmode \ \vskip -3cm \hskip 0.5cm
\epsfxsize=12cm\epsffile{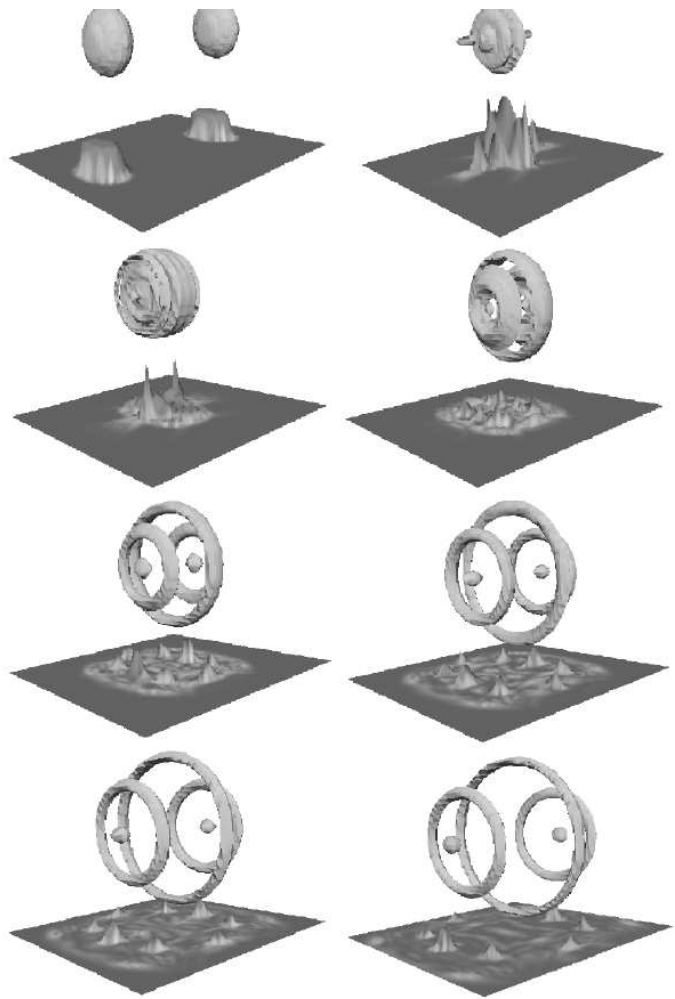}
\caption{The charge density at $t=0,27,33,39,45,60,75,90$ for two
Q-balls with $\omega_1=\omega_2=1.5$, initial positions $\pm(15,0,0)$,
velocities $v=0.8$ and relative phase $\alpha=0$. The high speed
collision leads to the formation of three loops, one at the point of
interaction, and two others which propagate back in opposite
directions along the line of interaction.}
\label{3dfast}
\end{center}
\end{figure}

Just to finish this section off and by way of illustrating that this
process is reasonably generic, we have performed an equivalent
simulation to figure~\ref{3d1} with a much higher speed of incidence
$(v=0.8)$. Due to the Lorentz contraction of the initial conditions,
this requires a smaller value of $\Delta x=0.15$ and consequently
$\Delta t=0.015$, and the results of this simulation are displayed in
figure~\ref{3dfast}. We see the production of a big loop at the point
of interaction plus two others which are repelled from the centre
along the line of interaction. At the end of the simulation the loops
are still expanding in size and are also getting close to the size of
the discrete grid. It is an interesting question as to whether loops
can be stable, and this question will be addressed in a separate
publication~\cite{BSut}.

\section{Q-ball Anti-Q-ball Dynamics}\news\ \ \ \ \ \

In the preceding sections we have studied in detail  2-Q-ball
interactions. The charge $Q$, however,  can also be negative; this
being achieved in the Q-ball solution by replacing $\omega$ with
$-\omega$ and these solutions are known as anti-Q-balls.  In this
section we will study the interactions of Q-ball/anti-Q-ball pairs  in
two and three dimensions.

Intuitively, one would expect slow soliton/anti-soliton interactions
with equal and opposite charges to result in annihilation into
radiation. However, only for a very small range of parameters does
this annihilation take place in the case of Q-balls due to the
complicated nature of the time-dependent interaction potential which
we have highlighted in the case of 2-Q-ball interactions.

In general a Q-ball/anti-Q-ball interaction will result in either the
two solitons bouncing back, or them passing through each other. In
both cases, the charge is partially annihilated, but only for a very
limited region of the interaction parameter space can it be thought of
as being complete. This absence of the annihilation can be attributed
to the concept of charge transfer which we have discussed  in 2-Q-ball
interactions\footnote{Here, as in the case of two Q-ball interactions
with different charges, the initial relative phase is a less important
concept. But charge transfer can still take place since the difference
between the two rotation speeds of the Q-balls is maximal.}  The main
difference between the charge transfer process for 2-Q-ball
interactions and the Q-ball/anti-Q-ball interactions under
consideration here is the two solitons now have opposite charge and
hence when charge is transfered it results in annihilation. However,
we have showed that the charge transfer never takes place fully in
2-Q-ball interactions and hence annihilation also never takes place
fully in a single interaction. When there is a lower bound on the
charge of a Q-ball it is more likely that sufficient charge transfer
can take place for complete annihilation, whereas in models where
arbitrarily small Q-balls can exist complete annihilation is likely to
be much more difficult.

\begin{figure}
\begin{center}
\leavevmode \ \vskip -3cm \hskip 0.5cm
 \epsfxsize=12cm\epsffile{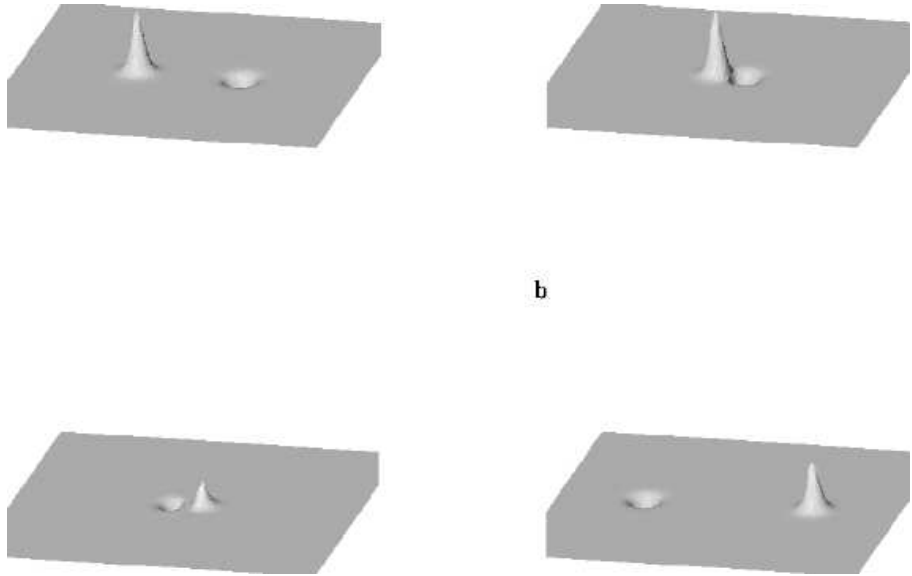}
\caption{Q-ball/anti-Q-ball interactions in two dimensions. The two
solitons are placed initially at $\pm(6,0)$, are Lorentz boosted
together with a velocity $v=0.3$ and have
$\omega_1=-\omega_2=1.8$. Partial annihilation of the Q-balls takes
place during the interaction and the momentum of the incident Q-ball is
sufficient to take it through to the other side.}
\label{e04}
\end{center}
\end{figure}

The process of partial annihilation, via charge transfer, is
illustrated in figure~\ref{e04} where we have displayed the charge
density at $t=0,15,20,50$ for the collision of a Q-ball and an
anti-Q-ball in two dimensions. The solitons were initially at
$\pm(6,0)$, with $\omega_1=-\omega_2=1.8$ and were Lorentz boosted
together with a velocity $v=0.3$. It is clear that the momentum of the
Q-ball carries it through the interaction region and that there is
some annihilation of the charge; the maximum charge density of the
outgoing Q-ball being lower than that for the incoming one. As we have
already discussed this is generically what takes place during a
Q-ball/anti-Q-ball collision. A variant on this kind of interaction is
that the Q-balls bounce back, again partially annihilating charge,
which takes place at low incident velocities for this particular
charge.

\begin{figure}
\begin{center}
\leavevmode \ \vskip -3cm \hskip 0.5cm
\epsfxsize=12cm\epsffile{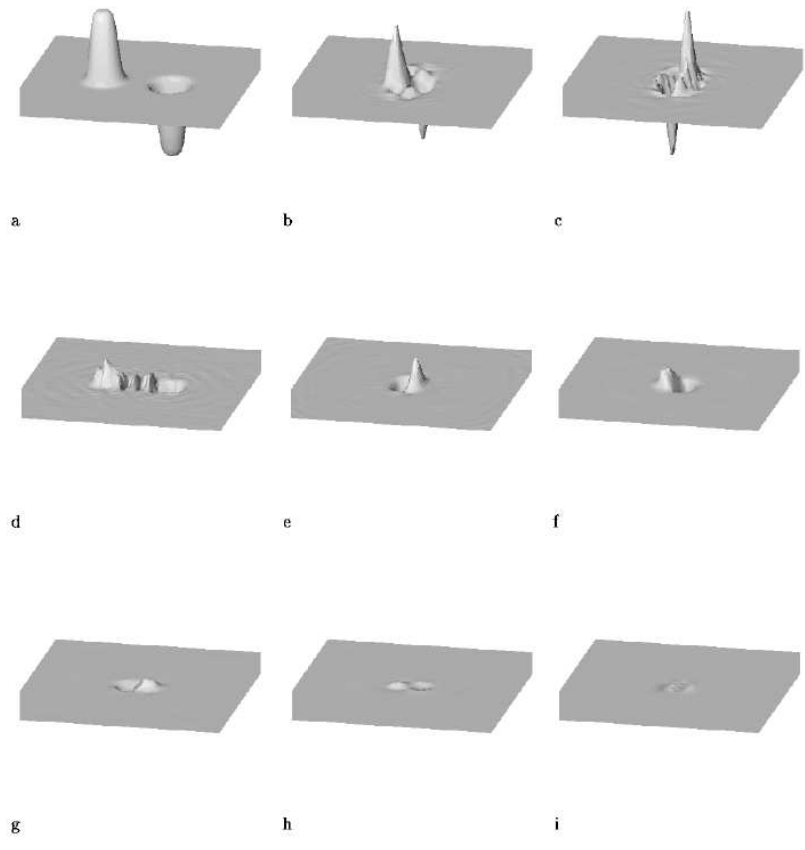}
\caption{As figure~\ref{e04} but the charge has been increased so that
$\omega_1=-\omega_2=1.5$. Complete annihilation takes place during a
complicated oscillatory interaction.}
\label{e03}
\end{center}
\end{figure}

\begin{figure}
\begin{center}
\leavevmode \ \vskip -3cm \hskip 0.5cm
\epsfxsize=12cm\epsffile{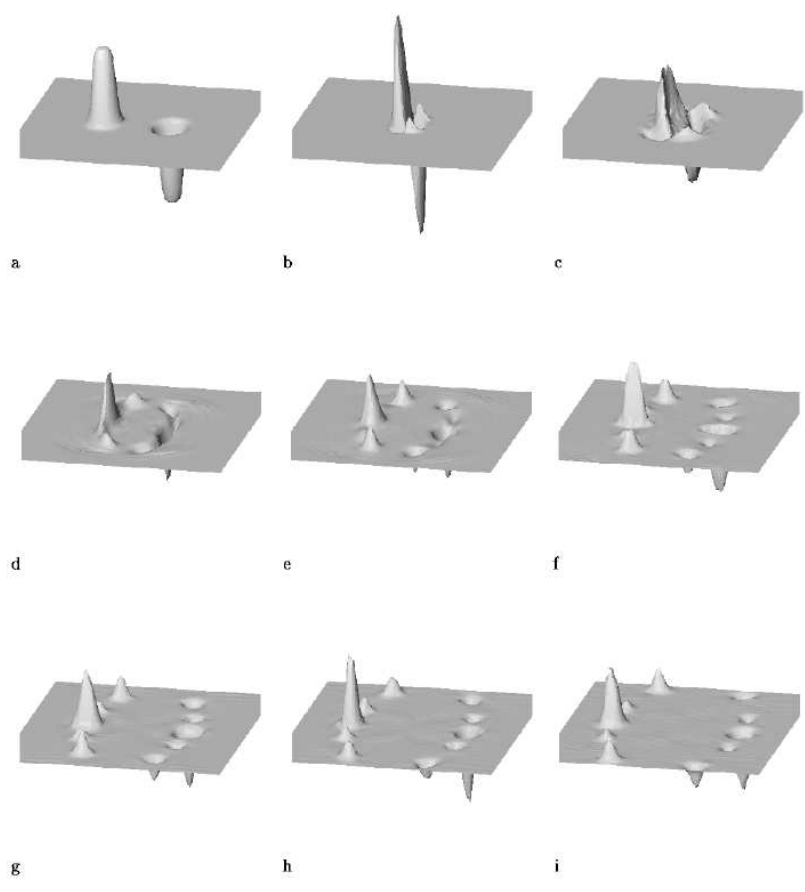}
\caption{As figure~\ref{e03} but with an incident velocity of
$v=0.6$. Notice that fission takes place in the plane perpendicular to
the line of incidence.}
\label{e02}
\end{center}
\end{figure}

This picture of partial annihilation with bounce back at low
velocities and the solitons passing through each other at high,
suggests that there exists some critical incident velocity $v_{\rm
c}$, a function of $\omega$ at which annihilation takes place, and
indeed this is what we find\footnote{In fact, there will exist a small
range of velocities around $v_{\rm c}$ for which complete annihilation
takes place.}. Figure~\ref{e03} illustrates this by displaying the
charge density at $t=0,25,50,100,150,250,300,350,400$ for a
Q-ball/anti-Q-ball collision, with the solitons initially positioned
at $\pm(6,0)$, charges $\omega_1=-\omega_2=1.5$, Lorentz boosted
together with velocity $v=0.3$. These are the same parameters as in
figure~\ref{e04}, except that the charge is much larger. It can be
clearly seen that annihilation eventually takes place, but even in
this case the mechanism is complicated, involving a number of
oscillations of the system before it is achieved. The charge here is
much higher than for the example above which had $\omega=1.8$, and in
the Q-ball interactions we saw complicated fission processes for
interactions involving solitons with this charge. This is also the
case in Q-ball/anti-Q-ball interactions as is illustrated in
figure~\ref{e02}, which uses the same parameters as in
figure~\ref{e03} except that the incident velocity is now much higher,
$v=0.6$.  The figures shown are at times
$t=0,10,15,25,30,35,40,45,50.$  One can see, after some partial
annihilation of charge, that the fission of the incident Q-balls takes
place in the direction perpendicular to the line of incidence and that
now there is an effective bounce back of the incident solitons with a
reduced charge.

\begin{figure}
\begin{center}
\leavevmode \ \vskip -3cm \hskip 0.5cm
\epsfxsize=12cm\epsffile{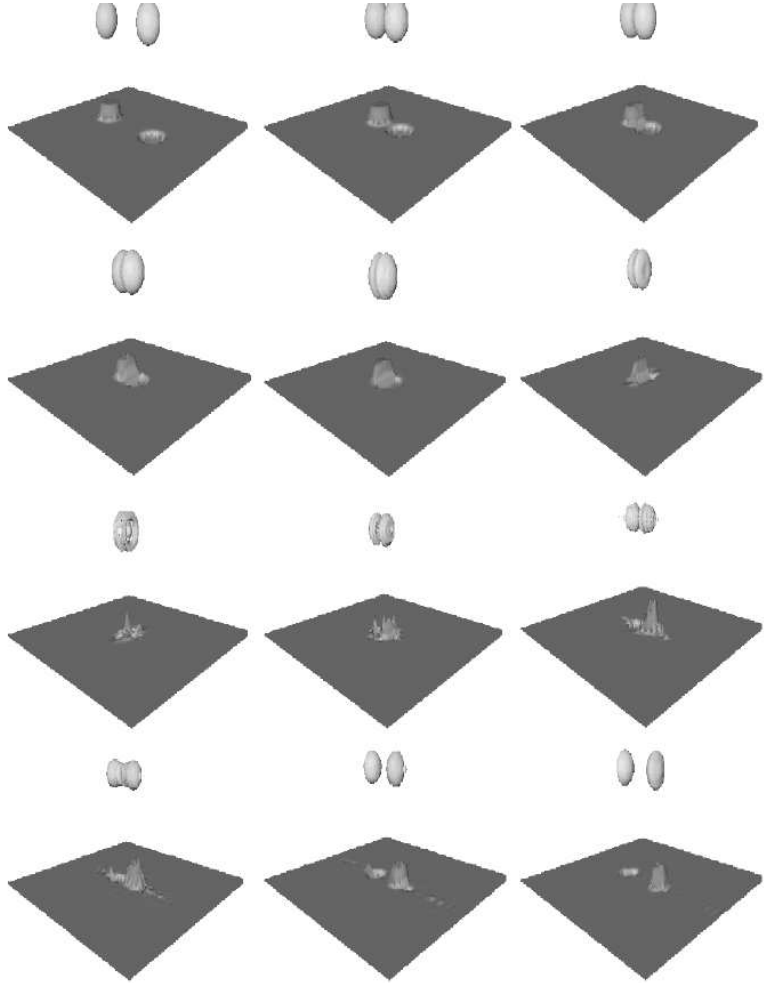}
\caption{Q-ball/anti-Q-ball interaction in three dimensions. Included
are isosurfaces of the modulus of the charge density and a slice of
the charge density itself through the centre of the interaction
region. The two solitons are initially at $\pm(10,0,0)$, Lorentz
boosted together with a velocity $v=0.3$ and have
$\omega_1=-\omega_2=1.5$. The two solitons pass through each other
with very little annihilation.}
\label{3danti3}
\end{center}
\end{figure}

\begin{figure}
\begin{center}
\leavevmode
\ \vskip -3cm \hskip 0.5cm
\epsfxsize=12cm\epsffile{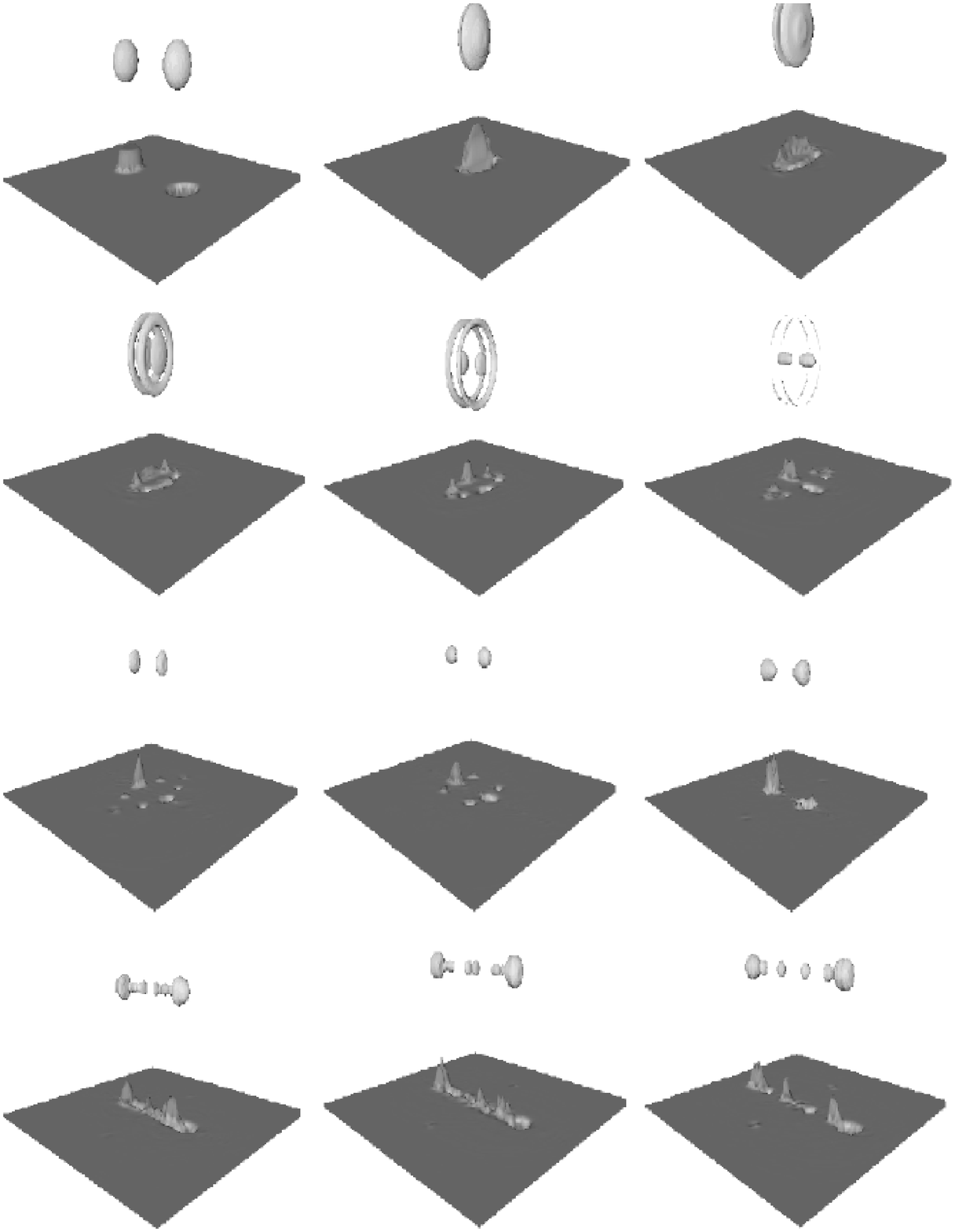}
\caption{As figure~\ref{3danti3} but with an incident velocity of
$v=0.6$. Two loops are formed in the initial interaction, which
subsequently collapse, leading to almost total annihilation.}
\label{3danti2}
\end{center}
\end{figure}

These processes are also prevalent in both one and three
dimensions. By way of illustration we have also included two examples
in three dimensions. In figure~\ref{3danti3}  the solitons initially
at $\pm(15,0,0)$, are Lorentz boosted together with a velocity $v=0.3$ and
have $\omega_1=-\omega_2=1.5$. In this particular interaction there
appears to be very little annihilation and the solitons effectively
pass through each other. Figure~\ref{3danti2} has the same initial
configuration, except that the solitons are boosted together with an
initial velocity of $v=0.6$. The subsequent interaction is complicated
involving first the formation of two loops, which is the equivalent of
the right-angled fission observed in two dimensions, and then what
appears to be almost total annihilation once the loops collapse. Two
small Q-balls are emitted along the line of incidence, along with much
radiation. 

\section{Summary and conclusions}\news\ \ \ \ \ \

We have identified the key parameters in the interactions of Q-balls to
be the relative phase, the incident velocity and the charge of Q-balls,
with the resulting interactions being strongly dependent on these parameters.
The generic interaction involves attraction if the relative phase
$\alpha=0$ and repulsion if $\alpha=\pi$ when the two Q-balls have the
same charge. If they have an initial relative phase other than
$\alpha=0$ or $\alpha=\pi$, or if the charges of the Q-balls are
different, the dynamics of the Q-balls results in 
charge transfer, a phenomena
which is analogous to that observed in discrete breather
systems (see, for example, ref.~\cite{ams} and references therein),
although this behaviour often has to be induced by making the solitons
come together via a Lorentz boost. With no Lorentz boost such breather
systems naturally repel and this is also seen after the charge
transfer process is complete. If the incident velocity is extremely
high Q-balls can be made to pass through each other with very little
interaction taking place.

This picture is almost independent of
the number of dimensions if the charge of the Q-balls is small (for
example, $\omega\approx 1.6$, but if the charge is much larger
fission can take place during the interaction process due to the
compression of the charge. In one dimension this process results in
the emission of Q-balls during the slow interaction of two large
Q-balls. In higher dimensions complicated, but analogous,
phenomena are observed in which fission takes place in the direction
perpendicular to the line of incidence. This leads, under specialized
circumstances to the right angled fission of Q-balls in two dimensions
and the production of loops in three. These fission effects can also
be coupled with those of attraction, repulsion and charge transfer to
produce complicated compound phenomena.

Interestingly, the naive expectation that an Q-ball/anti-Q-ball should
annhilate into radiation is modified by the complicated
breather-type interactions which take place. Since the difference
between the two charges is large this can be thought of a special case
of the charge transfer process, leading to the phenomenon of partial
charge annihilation, the case of complete annihilation being very
special. Fission of the Q-ball and anti-Q-ball can also take
place if the charge is high.

Our original motivation was to understand the
microphysical interactions of Q-balls formed by the Affleck-Dine
mechanism for baryogenesis within the framework of the MSSM. The
potential that we have concentrated on in this paper is different to
that expected in the MSSM, but as we have pointed out the main
interaction processes that we have identified are related to the
phase. Therefore, we expect our result to be qualitatively independent
of potential, and hence our results have some bearing on this
case. We have repeated several of the simulations described in this
paper for a potential of type II and found the same qualitative results.
The next step in this research is an analytic description of the
dynamics in terms of a slow manifold approach~\cite{BMS}, before their application to the
problem of Q-ball formation in the Early Universe.

\def\jnl#1#2#3#4#5#6{\hang{#1 [#2], {\it #4\/} {\bf #5}, #6.} }
\def\jnldot#1#2#3#4#5#6{\hang{#1 [#2], {\it #4\/} {\bf #5}, #6} }
\def\jnltwo#1#2#3#4#5#6#7#8{\hang{#1 [#2], {\it #4\/} {\bf #5},
#6;{\bf #7} #8.} }
\def\prep#1#2#3#4{\hang{#1 [#2],`#3', #4.} } 
\def\proc#1#2#3#4#5#6{{#1 [#2], in {\it #4\/}, #5, eds.\ (#6).} }
\def\book#1#2#3#4{\hang{#1 [#2], {\it #3\/} (#4).} }
\def\jnlerr#1#2#3#4#5#6#7#8{\hang{#1 [#2], {\it #4\/} {\bf #5}, #6.
{Erratum:} {\it #4\/} {\bf #7}, #8.} }
\def\prl{Phys.\ Rev.\ Lett.}
\def\pr{Phys.\ Rev.}
\def\pl{Phys.\ Lett.}
\def\np{Nucl.\ Phys.}
\def\prp{Phys.\ Rep.}
\def\rmp{Rev.\ Mod.\ Phys.}
\def\cmp{Comm.\ Math.\ Phys.}
\def\mpl{Mod.\ Phys.\ Lett.}
\def\apj{Ap.\ J.}
\def\apjl{Ap.\ J.\ Lett.}
\def\aap{Astron.\ Ap.}
\def\cqg{Class.\ Quant.\ Grav.} 
\def\grg{Gen.\ Rel.\ Grav.}
\def\mn{M.$\,$N.$\,$R.$\,$A.$\,$S.}
\def\ptp{Prog.\ Theor.\ Phys.}
\def\jetp{Sov.\ Phys.\ JETP}
\def\jetpl{Sov.\ Phys.\ JETP Lett.}
\def\jmp{J.\ Math.\ Phys.}
\def\zpc{Z.\ Phys.\ C}
\def\ijmp{Int.\ J.\ Mod.\ Phys.}
\def\cupress{Cambridge University Press}
\def\oup{Oxford University Press}
\def\pup{Princeton University Press}
\def\wss{World Scientific, Singapore}

\end{document}